\documentclass[twocolumn,usenames,dvipsnames]{aastex631}

\usepackage{graphicx}
\usepackage{bm}
\usepackage{amssymb,amsmath}
\usepackage{mathrsfs} 
\usepackage{latexsym}
\usepackage{mathtools} 
\usepackage{color}
\usepackage[normalem]{ulem} 
\usepackage{tikz}
\usepackage{pgfplots}
\usepackage{hyperref}
\usepackage{natbib}
\usepackage{multirow}
\usepackage{xspace}
\usepackage[nolist,nohyperlinks]{acronym}
\usepackage{booktabs}
\usepackage{enumitem}

\usepackage[deletedmarkup=xout,commentmarkup=uwave,commandnameprefix=ifneeded]{changes}
\usepackage{xcolor}
\definecolor{snsA}{HTML}{0173B2} 
\definecolor{snsB}{HTML}{DE8F05} 
\definecolor{snsC}{HTML}{029E73} 

\definechangesauthor[name=Max, color=teal]{MI}
\definechangesauthor[name=Aaron, color=purple]{AZ}
\definechangesauthor[name=Asad, color=brown]{AH}

\allowdisplaybreaks

\graphicspath{{.}}

\newcommand{\Austin}{\affiliation{Weinberg Institute, University of Texas at Austin, Austin, TX 78712, USA}}
\newcommand{\CCA}{\affiliation{Center for Computational Astrophysics, Flatiron Institute, NY}}
\newcommand{\CU}{\affiliation{
    Columbia University, Department of Astronomy, 550 West 120th Street, New York, NY, 10027, USA
}}

\newcommand{\R}[1]{\textcolor{red}{#1}}

\pgfplotsset{compat=1.18}

\makeatletter
\newcommand{\JSONLoadMacros}[2]{%
  \begingroup
    \def\JSONSetValue##1##2{\expandafter\gdef\csname json@\string#1@##1\endcsname{##2}}%
    \input{#2}%
  \endgroup}
\newcommand{\JSONParseValue}[2]{%
  \ifcsname json@\string#1@#2\endcsname
    \csname json@\string#1@#2\endcsname
  \else
    \@latex@warning{JSON value `#2' not found in \string#1}\textbf{??}%
  \fi}
\makeatother

\JSONLoadMacros{\FiducialMacros}{figures/FiducialModel/macros}
\JSONLoadMacros{\FourRegimeMacros}{figures/FourRegimes/macros}
\JSONLoadMacros{\BroadQMacros}{figures/broad_q/macros}

\begin{document}

\begin{acronym}
    \acro{GW}{gravitational wave}
    \acro{GR}{general relativity}
    \acro{CBC}{compact binary coalescence}
    \acro{BH}{black hole}
    \acro{NS}{neutron star}
    \acro{BBH}{binary black hole}
    \acro{LVK}{LIGO--Virgo--KAGRA}
    \acro{PE}{parameter estimation}
    \acro{FAR}{false-alarm rate}
    \acro{GWOSC}{the Gravitational Wave Open Science Center}
    \acro{SSB}{solar system barycenter}
    \acro{SPA}{stationary phase approximation}
    \acro{PN}{post-Newtonian}
    \acro{BNS}{binary neutron star}
    \acro{IMR}{inspiral-merger-ringdown}
    \acro{NSF}{National Science Foundation}
    \acro{GMM}{Gaussian mixture model}
    \acro{MCMC}{Markov chain Monte Carlo}
    \acro{MC}{Monte Carlo}
    \acro{BIC}{Bayesian information criterion}
	\acro{BF}{Bayes factor}
	\acro{SDDR}{Savage Dickey density ratio}
	\acro{TGMM}{truncated Gaussian mixture model}
	\acro{PPD}{population predictive distribution}
	\acro{HPDI}{highest posterior density interval}
	\acro{AGN}{active galactic nucleus}
	\acro{KDE}{kernel density estimate}
    \acro{CI}{credible interval}
    \acro{CL}{credible level}
    \acro{HBI}{hierarchical Bayesian inference}
    \acro{LVK}{LIGO-Virgo-KAGRA}
\end{acronym}

\def\checkmark{\tikz\fill[scale=0.4](0,.35) -- (.25,0) -- (1,.7) -- (.25,.15) -- cycle;}

\newcommand{\OneGTwoG}[0]{$1\mathrm{G}+2\mathrm{G}$}
\newcommand{\OneGOneG}[0]{$1\mathrm{G}+1\mathrm{G}$}
\newcommand{\TwoGTwoG}[0]{$2\mathrm{G}+2\mathrm{G}$}

\title{Three subpopulations of binary black holes in mass, spin magnitudes and tilts following GWTC-5}

\author[0000-0003-3491-5439]{Asad Hussain}
\CCA
\author[0000-0001-8830-8672]{Maximiliano Isi}
\CU
\CCA
\author[0000-0002-7453-6372]{Aaron Zimmerman}
\Austin
\date{\today}

\begin{abstract}
The mass and spin distributions of merging binary black holes offer clues to their formation pathways, which remain uncertain.
In this work, we model the population of binary black holes using generic subpopulations with distinct mass-ratio, spin-magnitude, and spin-tilt distributions, whose relative contributions vary with primary mass.
Using the 259 binary black holes in the fifth LIGO-Virgo-KAGRA catalog (GWTC-5), we resolve three subpopulations with low, medium, and high spins.
The low-spin population dominates for primary masses near $35\, M_\odot$, contributes roughly half of the binaries for primary masses of $10 \,M_\odot$, and favors nearly equal mass-ratios overall;
its spins are consistent with both black holes having negligible spin, with a log Bayes factor of $\log_{10}\mathcal{B}=\JSONParseValue{\FiducialMacros}{log_bayes_factors.in_favor_of_component_a_being_zero_spin}$ in favor of vanishing spins.
The medium-spin population accounts for the other half of the population around primary masses of $10\, M_\odot$ and favors spin magnitudes $\chi_i \approx 0.2$ for both components, preferentially aligned orientations, and comparable mass ratios.
The high-spin population dominates above primary masses of $60\, M_\odot$ and also contributes at lower masses ${\sim} 20\, M_\odot$; it favors a rapidly spinning primary ($\chi_1 \approx 0.7$), a broad range of secondary spins, widely distributed spin orientations, and a broad mass-ratio distribution.
Allowing for a fourth subpopulation yields no distinct additional mass-ratio or spin distribution, as it becomes degenerate with existing features.
To resolve highly concentrated features in spin, we control the variance of standard Monte Carlo likelihood estimators using truncated Gaussian mixture models.
Taken together, our results add to a growing body of evidence for substructure in the population of LIGO-Virgo-KAGRA black holes, providing further clues to uncover their formation histories.
\end{abstract}

\section{Introduction}

In the decade since the first detection of \acp{GW}, observations by Advanced LIGO~\citep{LIGOScientific:2014pky} and Advanced Virgo~\citep{VIRGO:2014yos} have produced hundreds of candidate signals from merging compact objects~\citep{LIGOScientific:2026sit,LIGOScientific:2026wfs,Nitz:2021zwj,Venumadhav:2019tad,Venumadhav:2019lyq,Zackay:2019btq,Olsen:2022pin,Mehta:2023zlk,Wadekar:2023gea}.
The vast majority of these have been \acp{BBH}, whose properties can reveal information about the lives and deaths of the massive stars that most likely produced them.
A central question is what channels form binaries that merge within a Hubble time.
There are multiple possibilities, including the isolated evolution of field binaries and the dynamical assembly of binaries in dense stellar environments \citep[e.g.,][]{Mandel:2018hfr,Mapelli:2021taw,Spera:2022byb}, which can include hierarchical mergers~\citep{Gerosa:2021mno}.
Each of these channels can leave distinct imprints on the observed \ac{BBH} population.

As the number of detected \acp{BBH} has grown, so has the evidence for subpopulations---loosely defined as groups of \acp{BBH} with distinct characteristics.
Multiple features also arise in the primary-mass distribution, most prominently a global peak near $10\,M_\odot$ and either another peak or a power-law break around $35\,M_\odot$~\citep[e.g.][]{LIGOScientific:2026ctl}.
\acp{BBH} also appear to split into a dominant low-spin subcomponent and a subdominant rapidly spinning subcomponent~\citep{Galaudage2021,Callister:2022qwb,Ray:2024hos,Hussain2025}.
Using GWTC-3, we previously identified these two spin subpopulations by jointly modeling the component-spin magnitudes~\citep{Hussain2025}.
Recent analyses additionally find that the spin and mass-ratio distributions vary with primary mass~\citep[e.g.,][]{Alvarez-Lopez:2026ymo,Flanagan:2026ayy,Ray:2026fdm,Tong:2025xir,Plunkett:2026pxt}, with hints of this mass dependence already present in earlier catalogs~\citep{2022ApJ...928...75H,Pierra:2024fbl,Berti:2025usa,Padhyegurjar:2026scg}.

Although the evidence for population structure is generally robust, the number and inferred properties of constituent subpopulations remain sensitive to modeling choices.
For instance, existing analyses characterize the mass dependence of the population in several complementary ways.
Some describe transitions or residual structure in the mass-ratio and effective-spin distributions as a function of primary mass~\citep{Alvarez-Lopez:2026ymo,Flanagan:2026ayy}, while flexible reconstructions of the joint distribution of primary mass, mass ratio $q$, effective inspiral spin $\chi_{\text{eff}}$ and effective precessing spin $\chi_p$ identify several mass-conditioned populations with distinct properties~\citep{Ray:2026uur,Ray:2026fdm,Cheng:2026bpc,Plunkett:2026pxt}.
Other analyses recover branches in the component-mass plane or separate low- and high-spin classes~\citep{Guttman:2026cnv,Li:2026iae,Zeeshan:2026mpr,Hussain:2026pfm}.

Here we search for \ac{BBH} subpopulations by applying a flexible population model to the latest \ac{LVK} catalog, GWTC-5~\citep{LIGOScientific:2026ctl}.
We allow multiple subpopulations to coexist across primary mass space, with distinct mass-ratio, spin-magnitude and spin-tilt distributions.
We find strong evidence for at least three \ac{BBH} subpopulations.
One component dominates across primary masses $m_1 \lesssim 50 \,M_\odot$, supports very low \ac{BH} spins, and favors nearly equal mass binaries.
The second component has a comparable fraction to the first at the lowest primary masses $m_1\lesssim 15 \, M_\odot$, beyond which it decreases to a negligible component.
Its spin distribution peaks at moderate dimensionless spin magnitudes of $\chi_i \sim 0.2$, and these spins are preferentially aligned with the orbital angular momentum.
The final component dominates at high \ac{BH} masses, features large dimensionless spins on the primary $\chi_1$ in $[0.4,1]$ (where we adopt 90\% credible bounds here and elsewhere), and broader but still preferentially large spins on the secondary, $\chi_2$ in $[0.2,1]$.
The distribution of spin orientations is broad and largely flat, consistent with isotropy but with some hints of structure. 

In our model, each subpopulation has a flexible mixing fraction that depends on primary mass, so that each can contribute at multiple mass scales and be constrained by events from across mass scales.
Each subpopulation is then distinguishable provided it has sufficiently distinctive features in mass-ratio and component-spin space.
However, most population analyses impose a finite smoothing scale or minimum feature width\footnote{An effective minimum width can also arise indirectly from a threshold on the variance of the estimated total log-likelihood, as commonly imposed in \ac{GW} population analyses~\citep[e.g.][]{LIGOScientific:2026ctl}.} that could blend neighboring features.
This is needed because \ac{MC} estimates of the hierarchical likelihood and selection function become increasingly noisy in the limit of narrow population features
~\citep{Farr:2019rap,Essick:2022ojx,Talbot:2023pex,Heinzel:2025ogf}.
We employ a method that maximizes the resolvability of narrow and closely spaced features in component-spin space by employing \ac{TGMM} estimators that are well-behaved in this limit~\citep{Hussain2025,Hussain:2025llf}.

While overall our findings are consistent with recent evidence for mass-dependent spin and mass-ratio structure~\citep[e.g.,][]{Alvarez-Lopez:2026ymo,Flanagan:2026ayy,Ray:2026fdm,Li:2026iae,Zeeshan:2026mpr,Cheng:2026bpc}, our analysis of GWTC-5 is novel in a number of ways:
\begin{enumerate}[label=(\alph*)]
    \item We characterize each subpopulation using the individual spin magnitudes and tilts of the binary components, without assuming the individual spins are identically distributed, rather than using effective spin parameters. This has the potential to leverage additional information from the catalog if it is present, and our results are readily interpretable in terms of astrophysical parameters.
    \item We explicitly model correlations between the cosine tilts $\cos\theta_1$ and $\cos\theta_2$ within every subpopulation.
    Recent work suggests that such correlations may retain information about binary formation that is absent from the individual tilt marginals~\citep{Dossena:2026cwf}.
    \item Our \ac{TGMM}-based approach allows us to explore population distributions that cannot be accessed with standard methods, in particular narrow population features where current methods fail to produce reliable likelihood estimates. Our approach reveals that the data are indeed consistent with narrow subpopulations in component-spin space, and in particular most of the \acp{BBH} likely have negligible spins.
\end{enumerate}
For example, applying our methods to GWTC-3~\citep{KAGRA:2021vkt} revealed the presence of a high-spin subpopulation in the two-dimensional space of spin magnitudes without the need for a mass-dependent spin model~\citep{Hussain2025}.
This high-spin subpopulation is robustly found by a number of other analyses using mass-dependent spin models, e.g.~in GWTC-3~\citep{Pierra:2024fbl}, GWTC-4~\citep{LIGOScientific:2025slb,Adamcewicz:2025phm,Guttman:2025jkv, Banagiri:2025dmy, Wolfe:2026meb, Cheng:2026bpc} and GWTC-5 \citep{Plunkett:2026pxt,Ray:2026fdm,Guttman:2026cnv}.

The remainder of this paper is organized as follows.
In Section~\ref{sec:Methods} we discuss our methods, models, and dataset.
In Section~\ref{sec:Results} we present our results and discuss each subpopulation in turn.
We conclude in Section~\ref{sec:Conclusions}.
The appendices provide further details of the model and additional results.

\section{Methods}
\label{sec:Methods}

\subsection{Model}
In order to study \ac{BBH} subpopulations, we perform hierarchical Bayesian inference \citep[e.g.,][]{Mandel:2018mve,Thrane:2018qnx,Vitale:2020aaz} to measure the hyperparameters $\boldsymbol{\Lambda}$ describing our population model $p(\boldsymbol{\theta} \mid \boldsymbol{\Lambda})$.
This gives the distribution of the intrinsic binary parameters, which we take to be the primary mass $m_1$, mass ratio $q = m_2/m_1 \leq 1$, redshift $z$ and dimensionless spin vectors $\boldsymbol{\chi}_1$, $\boldsymbol{\chi}_2$ of the primary and secondary.
We parametrize the spin vectors by their magnitudes ($\chi_1$, $\chi_2$) and the cosine tilts with respect to the orbital angular momentum ($\cos \theta_1$, $\cos \theta_2$), assuming that their azimuthal orientations are uniformly distributed.

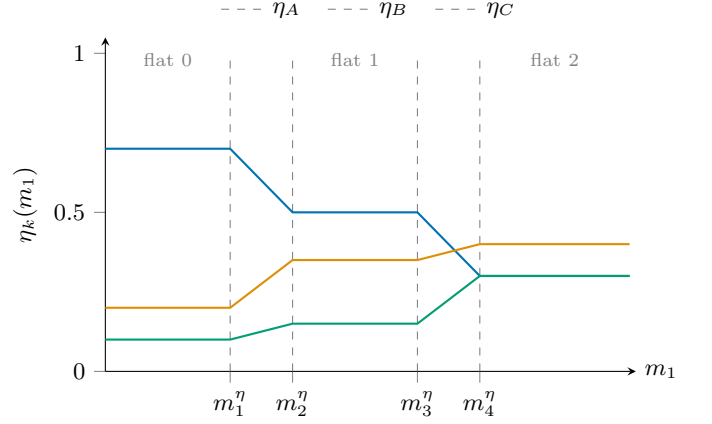
\begin{figure}[t]
    \centering
    \begin{tikzpicture}
    \begin{axis}[
        width=8.6cm, height=6cm,
        xmin=0, xmax=8.5,
        ymin=0, ymax=1.05,
        axis lines=left,
        xlabel={$m_1$},
        ylabel={$\eta_k(m_1)$},
        xtick={2,3,5,6},
        xticklabels={$m^\eta_1$,$m^\eta_2$,$m^\eta_3$,$m^\eta_4$},
        ytick={0,0.5,1},
        tick align=outside,
        every axis x label/.style={at={(current axis.right of origin)},anchor=west},
        legend style={
            at={(0.5,1.03)},
            anchor=south,
            legend columns=3,
            draw=none,
            /tikz/every even column/.append style={column sep=8pt}
        },
        clip=false,
    ]
    
    \addplot[dashed, gray] coordinates {(2,0) (2,1)};
    \addplot[dashed, gray] coordinates {(3,0) (3,1)};
    \addplot[dashed, gray] coordinates {(5,0) (5,1)};
    \addplot[dashed, gray] coordinates {(6,0) (6,1)};
    
    \addplot[thick, snsA] coordinates
        {(0,0.7) (2,0.7) (3,0.5) (5,0.5) (6,0.3) (8.4,0.3)};
    \addlegendentry{$\eta_A$}
    
    \addplot[thick, snsB] coordinates
        {(0,0.2) (2,0.2) (3,0.35) (5,0.35) (6,0.4) (8.4,0.4)};
    \addlegendentry{$\eta_B$}
    
    \addplot[thick, snsC] coordinates
        {(0,0.1) (2,0.1) (3,0.15) (5,0.15) (6,0.3) (8.4,0.3)};
    \addlegendentry{$\eta_C$}
    
    \node[gray, font=\scriptsize] at (axis cs:1,0.98)   {flat 0};
    \node[gray, font=\scriptsize] at (axis cs:4,0.98)   {flat 1};
    \node[gray, font=\scriptsize] at (axis cs:7.2,0.98) {flat 2};
    
    \end{axis}
    \end{tikzpicture}
    \caption{Schematic of the mass-dependent mixing fractions $\eta_k(m_1)$ for the three subpopulations $k\in\{A,B,C\}$.
    Each fraction is constant in three intervals of primary mass, connected by two linear transitions whose edges are the four shared, ordered break masses $m^\eta_1\leq\cdots\leq m^\eta_4$.
    The levels shown are illustrative; no ordering of the component labels is imposed during sampling, and the labels are assigned in post-processing as described in Appendix~\ref{appx:model}.}
    \label{fig:eta_cartoon}
\end{figure}

We use a single global model for the distribution of $m_1$ and $z$, but support three distinct subpopulations $k$ in $\{A, B, C\}$ with independent $\boldsymbol{\chi}_1$, $\boldsymbol{\chi}_2$ and $q$, with mixing fractions $\eta_k(m_1\mid\boldsymbol{\Lambda}^k_\eta)$ that depend on the primary mass.
More specifically, our model is
\begin{align}
    \label{eq:fullmodel}
    p(\boldsymbol{\theta} \mid \boldsymbol{\Lambda})  &=  
    p(z\mid\boldsymbol{\Lambda}_z)\, 
    p(m_1\mid\boldsymbol{\Lambda}_m)\, \times
    \notag \\
    &\sum_{k}
    \eta_k(m_1\mid\boldsymbol{\Lambda}_{\eta}^k)\, p_k(q,\boldsymbol{\chi}_1,\boldsymbol{\chi}_2\mid\boldsymbol{\Lambda}_\chi^k)\,,
\end{align}
where $p(z\mid\boldsymbol{\Lambda}_z)$ and $p(m_1\mid\boldsymbol{\Lambda}_m)$ are the fiducial, strongly-modeled distributions from \cite{LIGOScientific:2025pvj,LIGOScientific:2026ctl}: for the redshift $z$, a distribution uniform in comoving volume and time modified by a simple power law for $(1+z)$; for the primary mass $m_1$, a tapered broken power law plus two Gaussian peaks.
We model the mass-dependent mixing fractions $\eta_k(m_1\mid\boldsymbol{\Lambda}^k_\eta)$ such that they are constant within three tunable intervals of $m_1$, interpolating linearly between each interval as illustrated in Figure~\ref{fig:eta_cartoon}; the mixing fractions always add up to one for any given $m_1$, and the transition boundaries $m^\eta_1,\, m^\eta_2,\, m^\eta_3,\, m^\eta_4$ and level heights are inferred from the data.
This allows different subpopulations to dominate in different windows of primary mass, associating them with specific mass features if the data support this.

Within each subpopulation, the mass-ratio and spins are modeled using truncated normals, specifically 
\begin{align}
    p_k(q,\chi_1, \chi_2\mid\boldsymbol{\Lambda}^k) &= 
    \mathcal{N}_{[0,1]}(q\mid\mu_q^k,\sigma_q^k) \nonumber\\ &\times 
    \mathcal{N}_{[\boldsymbol{0},\boldsymbol{1}]}(\chi_1,\chi_2\mid\boldsymbol{\mu}_\chi^k,\boldsymbol{\Sigma}_\chi^k) \,,
\end{align}
and the tilts are modeled as a mixture of a free truncated normal and a component uniform in both tilts,
\begin{align}
     p_k(&\cos\theta_1, \cos\theta_2 \mid \boldsymbol{\Lambda_\theta}^k) = \nonumber \\
     & \left(1-\zeta_{\cos \theta}^k\right) U_{[-1,1]}(\cos\theta_1)\, U_{[-1,1]}(\cos\theta_2)
     \notag \\
      & 
     + \zeta_{\cos\theta}^k \mathcal{N}_{[-\boldsymbol{1},\boldsymbol{1}]}(\cos \theta_1,\cos \theta_2\mid\boldsymbol{\mu}_\theta^k,\boldsymbol{\Sigma}^k_\theta)\,.
\end{align}
Our fiducial model supports correlations among the two cosine tilts within each subpopulation, but does not allow correlations between the spin magnitudes (i.e.~$\boldsymbol{\Sigma}_\chi^k$ is diagonal).
We have also explored allowing for correlations between spin magnitudes, which results in nearly identical results at the cost of much slower sampling, and so we omit these correlations from the fiducial model.
Even though we do not allow for off-diagonal components in $\boldsymbol{\Sigma}_\chi^k$, correlated structure can still be identified in the population distribution after marginalizing over hyperparameters (see Sec.~\ref{sec:ppd} and our results below).

We do not apply any priors or restrictions on the labels of the three components.
This creates the possibility that during sampling the properties of distinct populations can swap, particularly if there is significant overlap in hyperparameter space between populations.
In practice this is not the case, and we instead find three distinct populations occupying different regions of hyperparameter space. 
For each sample we then sort the three components in ascending order of $\mu^k_{\chi_1} + \mu^k_{\chi_2}$, and relabel these $A,B,C$.
Further details of our fiducial population model and priors are given in Appendix~\ref{appx:Methods}.

We have explored the dependence of our results on our analysis choices by varying our model in a number of ways.
We find our results stable across these changes.
For example, our results are robust to enforcing the labeling based on the mass regions where we anticipate distinct components to dominate, although this slows the sampling significantly.

\subsection{Data and sampling}
Following \citet{LIGOScientific:2026ctl}, we use the 259 \acp{BBH} with false alarm rates $< 1\,\mathrm{yr}^{-1}$ from GWTC-5~\citep{LIGOScientific:2026sit,LIGOScientific:2026wfs} for population inference.
For these events, we use posterior samples inferred with the preferred waveform model or combinations of models as outlined in~\cite{LIGOScientific:2026ctl}.
In order to account for selection effects, we use the results of large scale injection-recovery campaigns and semi-analytic sensitivity estimates~\citep{ligo_scientific_collaboration_2026_19500052} 
to compute the detection fraction $\xi(\boldsymbol{\Lambda})$ for a given population model,
as done in~\cite{LIGOScientific:2025pvj,LIGOScientific:2026ctl}.
The injection-recovery campaigns and semi-analytic sensitivity estimates are described in~\cite{KAGRA:2021duu,Essick:2023toz,Essick:2025zed}.

To control the variance in standard \ac{MC} estimation of 
the hierarchical likelihood and detection fraction, we fit our event posterior samples and the results of the injection-recovery campaigns
to \acp{TGMM} as in \citet{Hussain2025,Hussain:2026pfm}.
Using these, our methods provide reliable estimates for the likelihood and detection fraction even when the population model is concentrated in one or more dimensions~\citep{Hussain:2025llf}.  
Our techniques are critical for our study of \ac{BBH} subpopulations, since we find that one subpopulation is highly concentrated at small spin magnitudes.

We use the \textsc{TruncatedGaussianMixtures} package~\citep{truncatedgaussianmixtures} to perform our \ac{TGMM} fits.
We perform our Bayesian analysis to estimate the hyperposterior using the \textsc{gravpop} package~\citep{gravpop} and sample the hyper-posteriors using the no-U-turn sampler \citep{2011arXiv1111.4246H} Hamiltonian \ac{MC} algorithm \citep{2011hmcm.book..113N,2017arXiv170102434B} as implemented in \textsc{numpyro}~\citep{phan2019composable,bingham2019pyro}.
We compute \acp{BF} from the posterior using the Savage-Dickey density ratio~\citep{dickey1971weighted,Verdinelli01061995}.

\subsection{Population predictive distributions} \label{sec:ppd}

Besides reporting our inferred hyperparameter posteriors, we also produce plots of \acp{PPD}.
For binary parameters $\boldsymbol{\theta}$, the \ac{PPD} is the posterior distribution of $\boldsymbol{\theta}$ given the catalog data $\mathcal D$, marginalized over the hyperparameters $\boldsymbol{\Lambda}$, namely
\begin{align} \label{eq:PPD}
    p(\boldsymbol{\theta}\mid \mathcal D) = \int p(\boldsymbol{\theta}\mid\boldsymbol{\Lambda})\, p(\boldsymbol{\Lambda}\mid \mathcal D)\, {\rm d}\boldsymbol{\Lambda}\,.
\end{align}
We compute this by sampling the hyperparameters $\boldsymbol{\Lambda}$ from their hyperposteriors $p(\boldsymbol{\Lambda}\mid \mathcal D)$ and then evaluating the population model of $\boldsymbol{\theta}$ for each sample.
When considering two parameters at a time, we take draws of $\boldsymbol{\theta}$ from the population model $p(\boldsymbol{\theta}\mid\boldsymbol{\Lambda})$ for each hyperparameter sample; this results in an estimator for the mean population in two dimensions.
When considering a single parameter $\theta$, we instead evaluate the population model at each $\theta$ value conditioned on the hyperposterior sample, then find the median and $90\%$ \ac{CI} over the hyperposterior draws at each $\theta$ value, and plot these as our estimate of the \ac{PPD} and its uncertainty. 

The same procedure can be defined for individual subpopulations by conditioning on the label $k$ in $\{A,B,C\}$, namely
\begin{align} \label{eq:PPD_subpopulation}
    p(\boldsymbol{\theta}\mid \mathcal D, k) = \int p(\boldsymbol{\theta}\mid\boldsymbol{\Lambda}, k)\, p(\boldsymbol{\Lambda}\mid \mathcal D, k)\, {\rm d}\boldsymbol{\Lambda}\,.
\end{align}
The result loses information about the relative weights of the subpopulations, but this can be recovered from the posteriors on the fractions $\eta_k(m_1)$.

\section{Results}
\label{sec:Results}

\begin{figure*}
    \centering
    \includegraphics[width=\columnwidth]{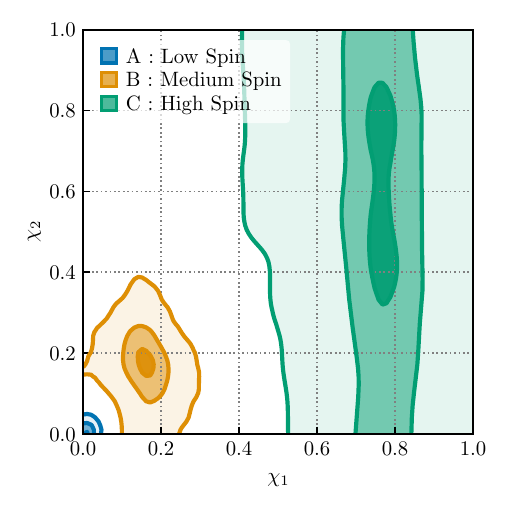}
    \includegraphics[width=\columnwidth, height=\columnwidth]{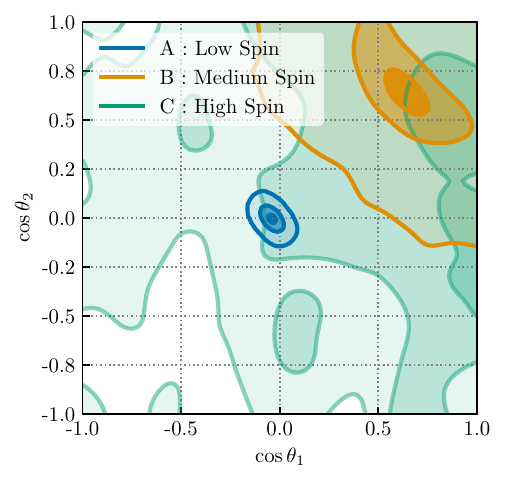}
    \caption{Inferred \acp{PPD} for the spin magnitudes (left) and cosine tilts (right) of the three subpopulations, with contours enclosing the 10\%, 50\% and 90\% credible regions. 
    One population is highly concentrated towards very low spins, with spin components along the orbital angular momentum disfavored.  
    The second population prefers spin magnitudes $\chi_i \sim 0.2$ and spin tilts preferentially aligned with the orbital angular momentum. The third population prefers high primary spin magnitudes $\chi_1 \gtrsim 0.5$, a wide distribution of secondary spin magnitudes, and a wide distribution of tilt angles.}
    \label{fig:SpinPPDs}
\end{figure*}

\begin{figure}[tb]
    \centering
    \includegraphics[width=\columnwidth]{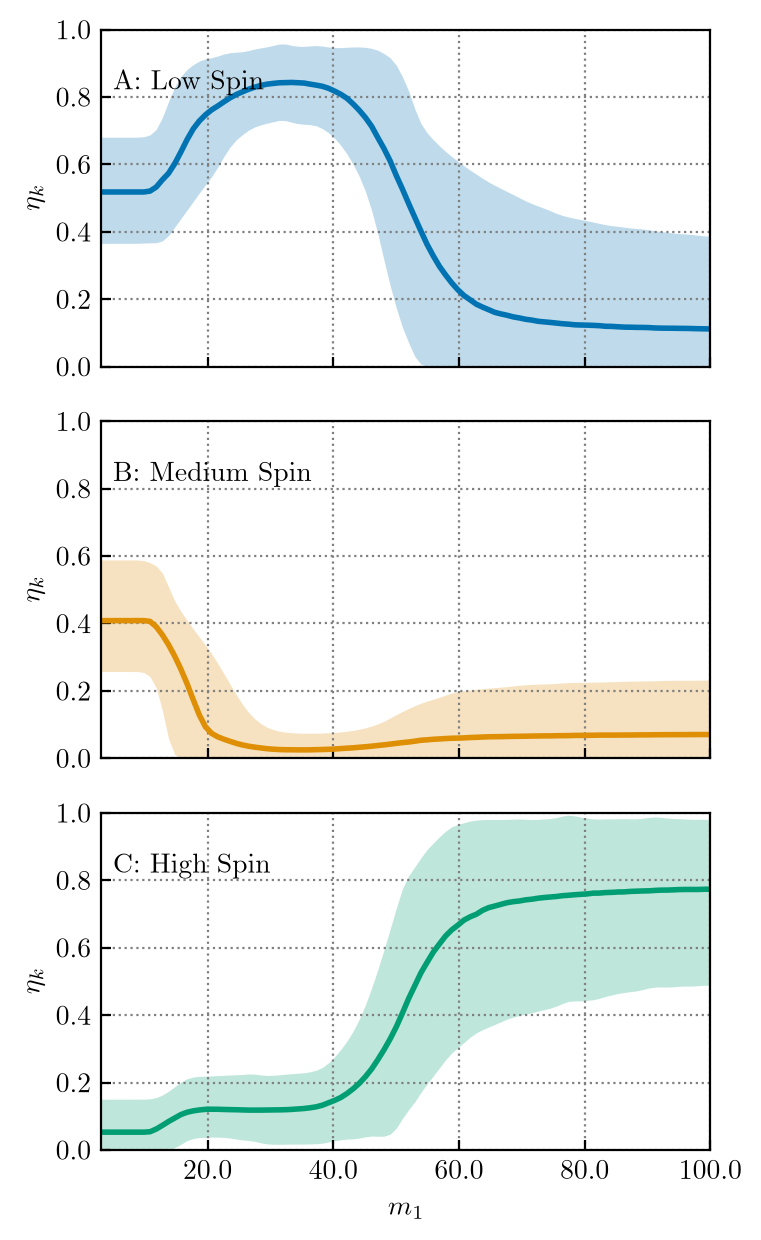}
    \caption{Inferred distributions for the mixing fractions $\eta_k(m_1)$ among the three populations, constrained to sum to unity. The three populations are labeled according to their distinct spin distributions (Figure~\ref{fig:SpinPPDs}). 
    The low-spin population dominates for $m_1 \lesssim 60 \, M_\odot$, especially near $m_1 \sim 35 \, M_\odot$, while the medium-spin population has a comparable size for $m_1 \lesssim 15 M_\odot$. The high-spin component dominates at high masses $m_1 \gtrsim 60 \, M_\odot$, with small but nonzero contribution at $m_1 \sim 20\, M_\odot$ extending towards higher masses. }
    \label{fig:fraction}
\end{figure}

\begin{figure}[tb]
    \centering
    \includegraphics[width=\columnwidth]{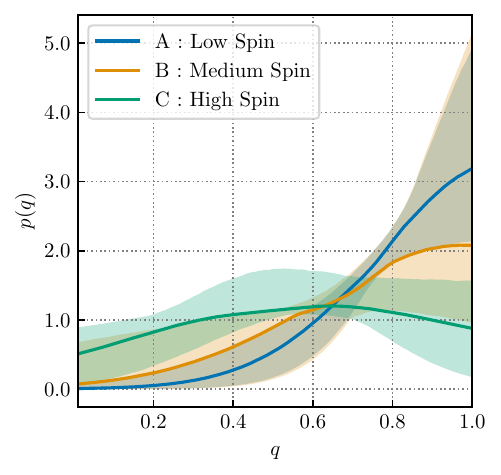}
    \caption{Inferred mass-ratio $q$ distributions for the three subpopulations. The low-spin population shows the strongest preference for nearly equal-mass binaries, while the high-spin population is broad and nearly flat.}
    \label{fig:MassRatio}
\end{figure}

Applying our model to the \acp{BBH} in GWTC-5, we find three subpopulations with distinct spin and mass-ratio distributions.
Figure~\ref{fig:SpinPPDs} shows our inferred \acp{PPD} for spin magnitudes and cosine tilts.
The three populations fall neatly into low-spin (subpopulation A), medium-spin (B), and high-spin (C) categories.
The corresponding \acp{PPD} of the mixing fractions $\eta_k(m_1)$ are illustrated in Figure~\ref{fig:fraction}.
We see that the low-spin population is dominant at low and intermediate values of $m_1$, the medium-spin population is significant at the lowest masses, and the high-spin component dominates strongly at the highest masses, with nonzero support around $m_1 \sim 20\, M_\odot$.
The mass-ratio \acp{PPD} are shown in Figure~\ref{fig:MassRatio}, where the low- and medium-spin populations peak towards equal mass binaries, while the high-spin population has a nearly flat distribution.

In the following we discuss each of these three subpopulations in more detail. We then consider the possibility of a fourth subpopulation.

\subsection{Low-spin population}
\label{sec:low_spin}

The low-spin population is highly concentrated towards negligible spins, with location parameters ${\boldsymbol \mu}_\chi^A = (\JSONParseValue{\FiducialMacros}{subpop_hyperposteriors.a.mu.chi_1}, \JSONParseValue{\FiducialMacros}{subpop_hyperposteriors.a.mu.chi_2})$ and widths ${\boldsymbol \sigma}_{\chi}^A = (\JSONParseValue{\FiducialMacros}{subpop_hyperposteriors.a.sigma.chi_1}, \JSONParseValue{\FiducialMacros}{subpop_hyperposteriors.a.sigma.chi_2})$.
It is consistent with a population of \acp{BH} with negligible spins: we find the \ac{BF} in favor of zero spins to be $\log_{10}\mathcal{B}=\JSONParseValue{\FiducialMacros}{log_bayes_factors.in_favor_of_component_a_being_zero_spin}$.
This subpopulation marginally dominates at low masses, with $\eta_A(10 M_\odot) = \JSONParseValue{\FiducialMacros}{fraction_at_masses.fraction_at_mass_10.A}$ 
below the first inferred mass break, which occurs at 
$m_1^\eta = \JSONParseValue{\FiducialMacros}{mass_breakpoints[0]} M_\odot$.
It is strongly dominant in an intermediate mass range $\JSONParseValue{\FiducialMacros}{mass_breakpoints[1]} M_\odot 
< m_1 < 
\JSONParseValue{\FiducialMacros}{mass_breakpoints[2]} M_\odot$,
with 
$\eta_A(35 M_\odot) =  \JSONParseValue{\FiducialMacros}{fraction_at_masses.fraction_at_mass_35.A}$, 
and its fraction decreases and becomes more uncertain at high masses.
Among the three subcomponents, the low-spin population shows the greatest preference for nearly equal-mass systems, with
$\mu^A_q = \JSONParseValue{\FiducialMacros}{subpop_hyperposteriors.a.mu.mass_ratio}$ and $\sigma^A_q = \JSONParseValue{\FiducialMacros}{subpop_hyperposteriors.a.sigma.mass_ratio}$.
\footnote{
Note that our prior ranges on the truncated-normal parameters extend outside the physical domain, allowing for a greater range of functional shapes. See Appendix~\ref{appx:Methods}} 

Given its dominance at moderate masses and substantial inferred fraction at low masses, we infer that most of the \acp{BH} observed with \acp{GW} lie in this population.
In Appendix~\ref{appx:FurtherResults} we give the assignment probabilities of each source in GWTC-5 to each of the three subpopulations.
This population hosts the ``typical'' \ac{LVK} \acp{BBH}, such as GW150914 and GW250114, with low spins and masses around $35\,M\odot$.
The large number of observed events in this dataset helps to explain how we can arrive at such tight constraints on the spin distribution: from the \acp{PPD}, we find $\chi_1 = \JSONParseValue{\FiducialMacros}{subpop_ppds.a.chi_1}$
and $\chi_2 =  \JSONParseValue{\FiducialMacros}{subpop_ppds.a.chi_2}$,
so that the 90\% credible bounds on these are $\lesssim 0.06$.
With $N_A \sim 100$ events observed in this population, such bounds can arise even from the loose constraints available on $\chi_1$ and $\chi_2$ from typical events ($\sigma_\chi \sim 1$ and $1/\sqrt{N_A} \lesssim 0.1$).

To whatever extent the spin magnitudes are not zero in this subpopulation, their tilts must be concentrated near the equatorial plane. 
It has been noted previously that the tilt distribution can be strongly influenced by the population model and prior~\citep{Vitale:2022dpa,Vitale:2025lms,Wolfe:2026meb}.
While effects such as three-body interactions in triple systems have the potential for generating spin tilts preferentially in the orbital plane~\citep{Antonini:2017ash,Liu:2018nrf,Rodriguez:2018jqu,Stegmann:2025zkb}, we attribute our result to the fact that the data more strongly constrain spin components aligned with the orbital angular momentum, $\chi_i \cos \theta_i$, than spin components in the plane  $\chi_i \sin \theta_i$. 
For a population with negligible spins, we expect \ac{GW} measurements to allow a greater range of spin magnitudes in the equatorial plane than along the orbital angular momentum.
This fact, together with our nearly uniform prior \acp{PPD} for the spin-tilts, leaves most of the allowed posterior density in the plane.
To underscore this point, in Appendix~\ref{appx:Simulation} we carry out population inference on a toy model of a simulated binary population with negligible spins, assuming the spin posterior of each event is shaped similarly to our expectations for real GW events, and find a qualitatively similar result as in Figure~\ref{fig:SpinPPDs}.
This effect has also been recently reported as an expected feature when modeling the global spin population with component spins~\citep{Wolfe:2026pjn}. 

The existence of a subpopulation with small or negligible spins is consistent with other studies \citep[e.g.,][]{Galaudage2021,Ray:2024hos,Banagiri:2025dmy,Cheng:2026bpc,Flanagan:2026btu,Guttman:2026cnv,Alvarez-Lopez:2026ymo,Li:2026iae,Zeeshan:2026mpr}, including our previous analysis of GWTC-4 with a two-component model~\citep{Hussain:2026pfm}.
Black holes with negligible spins can arise from a stellar population where efficient mechanisms transport angular momentum from a star's core to its envelope during isolated stellar evolution~\citep{Fuller:2019ckz,Ma:2019cpr,Fuller:2019sxi,Belczynski:2017gds}.
This population is broadly consistent with the isolated binary evolution channel in cases where angular momentum transport is efficient and subsequent binary interactions do not spin up one of the stellar cores after the envelope is ejected. 
It is also consistent with dynamically formed binaries, where each \ac{BH} evolved in isolation and masses were correlated through mass segregation~\citep{Rodriguez:2016avt,Park:2017zgj,Antonini:2022vib,Chattopadhyay:2022buz}.

\subsection{Medium-spin population}
\label{sec:medium_spin}

The medium-spin population has moderate spin magnitudes with
$\chi_1 = \JSONParseValue{\FiducialMacros}{subpop_ppds.b.chi_1}$ and
$\chi_2 = \JSONParseValue{\FiducialMacros}{subpop_ppds.b.chi_2}$ based on the \ac{PPD}, thus both spins are statistically favored to be away from zero.
The tilts are preferentially aligned with the orbital angular momentum, with $\cos \theta_1 = \JSONParseValue{\FiducialMacros}{subpop_ppds.b.cos_tilt_1}$ and
$\cos \theta_2 = \JSONParseValue{\FiducialMacros}{subpop_ppds.b.cos_tilt_2}$. 
We infer 
${\boldsymbol \mu}_\chi^B = 
(\JSONParseValue{\FiducialMacros}{subpop_hyperposteriors.b.mu.chi_1}, 
\JSONParseValue{\FiducialMacros}{subpop_hyperposteriors.b.mu.chi_2})$ 
and 
${\boldsymbol \sigma}_{\chi}^B 
= (\JSONParseValue{\FiducialMacros}{subpop_hyperposteriors.b.sigma.chi_1}, 
\JSONParseValue{\FiducialMacros}{subpop_hyperposteriors.b.sigma.chi_2})$ for the spin-magnitude hyperparameters,
${\boldsymbol \mu}_{\cos \theta}^B 
= (\JSONParseValue{\FiducialMacros}{subpop_hyperposteriors.b.mu.cos_tilt_1}, 
\JSONParseValue{\FiducialMacros}{subpop_hyperposteriors.b.mu.cos_tilt_2})$ and
${\boldsymbol \sigma}_{\cos\theta}^B 
= (\JSONParseValue{\FiducialMacros}{subpop_hyperposteriors.b.sigma.cos_tilt_1}, 
\JSONParseValue{\FiducialMacros}{subpop_hyperposteriors.b.sigma.cos_tilt_2})$ for the spin-tilt hyperparameters, with a correlation parameter of $\rho^B_{\cos\theta} = \JSONParseValue{\FiducialMacros}{subpop_hyperposteriors.b.rho}$ for the latter.
From these we see that highly concentrated spin-magnitude and tilt distributions are allowed, but not required, by the data.

This population constitutes a significant fraction of the total at the lowest primary masses, below $m_1^\eta = \JSONParseValue{\FiducialMacros}{mass_breakpoints[0]} \, M_\odot$, with 
$\eta_B(10\, M_\odot) = \JSONParseValue{\FiducialMacros}{fraction_at_masses.fraction_at_mass_10.B}$.
Unlike for the low-spin population, this fraction decreases within the first transition regime of $m_1$ within $\JSONParseValue{\FiducialMacros}{mass_ranges.Transition[0]} \, M_\odot$, beyond which the inferred fraction is consistent with $\eta_B = 0$; this means that this medium-spin subpopulation only appears to be relevant for $m_1 \lesssim 20\,M_\odot$.
Like the low-spin population, it peaks towards equal masses, with 
$\mu^B_q = \JSONParseValue{\FiducialMacros}{subpop_hyperposteriors.b.mu.mass_ratio}$ 
and 
$\sigma^B_q = \JSONParseValue{\FiducialMacros}{subpop_hyperposteriors.b.sigma.mass_ratio}$.
It has somewhat more support towards unequal masses relative to the low-spin population.

The proximity of this population in mass and spin space to the low-spin population means that we expect the two to blend together when the data are analyzed with models that do not have sufficient flexibility.
We also expect that with smaller datasets the two may be hard to distinguish.
This third population was identified in GWTC-4 data by~\cite{Cheng:2026bpc} using a flexible reversible-jump-\ac{MCMC} approach.
When our own \ac{TGMM} methods are applied to GWTC-4 data, allowing for only two subpopulations, we previously found the low- and medium-spin populations blended into one covering both regimes ~\citep{Hussain:2026pfm}, with qualitatively similar results for the high-spin population as we find here.
When the \ac{TGMM} methods are applied to GWTC-5, allowing for two subpopulations but with less flexibility in the spin-tilt model (pinning the truncated normal to peak at $\cos \theta_i = 1$ as done in \citealt{KAGRA:2021duu}), the blend of low- and medium-spin populations mimics the properties of the medium-spin population~\citep{LIGOScientific:2026ctl}.
Other analyses of GWTC-5 also identify changing spin structure or an additional spinning contribution at low primary mass~\citep{Flanagan:2026btu,Ray:2026fdm,Alvarez-Lopez:2026ymo}.
However, analyses that assume a single spin distribution within a corresponding low mass region, such as those focusing on the $10\, M_\odot$ peak~\citep[e.g.,][]{Flanagan:2026btu,Alvarez-Lopez:2026ymo,Guttman:2026cnv} would inevitably mix the two populations we identify here.
 
This population is broadly compatible with isolated binary evolution in which stable mass transfer or tidal synchronization produces a modest, preferentially aligned spin on the second-born black hole~\citep{Qin:2018vaa,Bavera:2020inc,Olejak:2021iux}.
Mass-ratio reversal can additionally produce systems in which the initially less massive star becomes the more massive black hole while binary interactions spin up its companion~\citep{Broekgaarden:2022nst,Mould:2022xeu}.
However, mass-ratio reversal tends to predict that only one of the two \acp{BH} is spinning appreciably, whereas our model prefers a population where both \acp{BH} tend to have nonzero spin.
For example, the hyperposteriors (Appendix~\ref{appx:FurtherResults}) show that if the location parameter of the secondary spin $\mu^B_{\chi_2} \approx 0$, then the corresponding width is nonzero with $0.1 \lesssim \sigma^B_{\chi_2} \lesssim 0.15$; this means that at least some binaries must have $\chi_2 > 0$.
Meanwhile the probability that both location parameters simultaneously exceed $\mu^B_{\chi_i} = 0.1$ is $P = 0.8$, again favoring spins on both \acp{BH}.

\subsection{High-spin population}
\label{sec:high_spin}

\begin{figure}[tb]
    \includegraphics[width=\columnwidth]{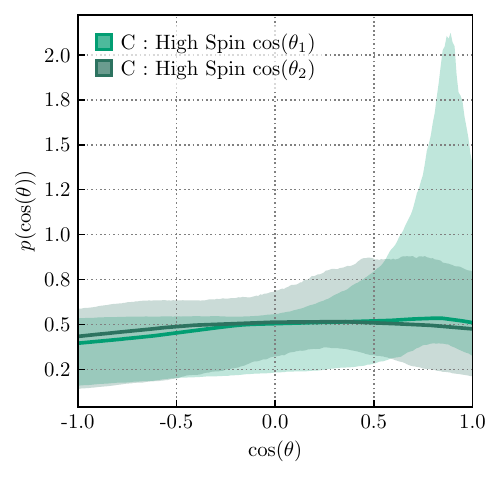}
    \caption{Marginal \acp{PPD} of $\cos\theta_1$ and $\cos\theta_2$ of the high-spin population. The distributions are fairly uniform, but populations where the spin tilt of the primary is preferentially aligned with the orbital angular momentum are allowed.}
    \label{fig:high_spin_tilt_marginals}
\end{figure}

\begin{figure}[tb]
    \includegraphics[width=\columnwidth]{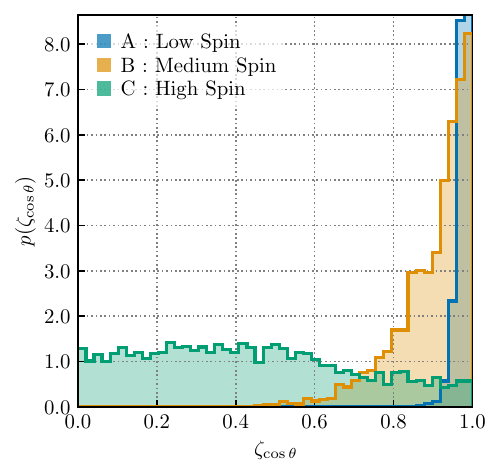}
    \caption{Posterior over the weight $\zeta_{\cos \theta}^k$ assigned to the truncated normal component of the spin-tilt model for each subpopulation. The distribution for the low-spin population extends beyond the plot range.}
    \label{fig:zeta_tilt_hyperposterior}
\end{figure}

The high spin population favors rapid spins on the primary, with $\chi_1 = \JSONParseValue{\FiducialMacros}{subpop_ppds.c.chi_1}$ and a secondary spin magnitude more widely distributed but with support at high secondary spins as well, $\chi_2 = \JSONParseValue{\FiducialMacros}{subpop_ppds.c.chi_2}$, both based on the \ac{PPD}.
This population has widely distributed spin tilts, although
the tilt \ac{PPD} is not completely uniform and disfavors scenarios where both \ac{BH} spins lie below the equatorial plane (Figure~\ref{fig:SpinPPDs}).
The 1D marginal \acp{PPD} shown in Figure~\ref{fig:high_spin_tilt_marginals} hide this correlation; overall they are relatively flat, but demonstrate that populations where $\theta_1$ is preferentially small are supported by the data.
Our inference on the mixing fractions $\zeta_{\cos \theta}^k$ between isotropic spin tilts and a truncated normal for all three populations is shown in Figure~\ref{fig:zeta_tilt_hyperposterior}.
We see that the low- and medium-spin populations strongly disfavor the isotropic spin hypothesis, but the high-spin population is consistent with the purely isotropic case, $\zeta_{\cos \theta}^C = 0$.

The high-spin population has support in two mass ranges.
The first is a range of low masses above the first break $m^\eta_1 = \JSONParseValue{\FiducialMacros}{mass_breakpoints[0]} \, M_\odot$ and below the third $m^\eta_3 = \JSONParseValue{\FiducialMacros}{mass_breakpoints[2]}\, M_\odot$. 
It is most prominent near the second break that marks the end of the first transition region at $m^\eta_2 = \JSONParseValue{\FiducialMacros}{mass_breakpoints[1]}\, M_\odot$,
where it makes a relatively small but nonzero fraction 
$\eta_C(20\, M_\odot) = \JSONParseValue{\FiducialMacros}{fraction_at_masses.fraction_at_mass_20.C}$. 
The second is at high masses 
$m_1 \JSONParseValue{\FiducialMacros}{mass_ranges.Flat[2]}\, M_\odot$, where it dominates over the other two subpopulations with 
$\eta_C(60\, M_\odot) = \JSONParseValue{\FiducialMacros}{fraction_at_masses.fraction_at_mass_60.C}$, and may make up essentially all of the \acp{BBH}.
In the intermediate mass range where the low-spin population dominates, the fraction decreases but may be nonzero, with an inferred 
$\eta_C(35 \, M_\odot) = \JSONParseValue{\FiducialMacros}{fraction_at_masses.fraction_at_mass_35.C}$.
The mass-ratio distribution is nearly flat, showing the widest support among the three and a hint of a peak near mass ratios $\sim 1:2$, with
$\mu^C_q = \JSONParseValue{\FiducialMacros}{subpop_hyperposteriors.c.mu.mass_ratio}$ and $\sigma^C_q = \JSONParseValue{\FiducialMacros}{subpop_hyperposteriors.c.sigma.mass_ratio}$.
We plot our inferred hyperposteriors and \ac{PPD} for the effective spin parameters $\chi_\mathrm{eff}$ and $\chi_p$ in Appendix~\ref{appx:FurtherResults}.

For our fiducial model, we restrict the prior range on the mass-ratio hyperparameters $\mu_q$ and $\sigma_q$, for example requiring $\sigma_q \geq 0.2$. 
While our TGMM methods control the variance in the likelihood in the spin sector, where we can integrate analytically, in the mass sector we approximate integrals via \ac{MC} averaging; as a consequence, narrow mass ratio distributions lead to higher variance in our likelihood estimates, so the prior disallows that.
As a control, in Appendix~\ref{appx:FurtherResults} we relax these priors at the cost of greater variance.
When doing so, the mass ratio \ac{PPD} for the high spin population remains broad and relatively flat, but with greater weight towards lower $q$ values. 
We also find mild confusion between the high- and medium-spin populations in some regions of parameter space, but broad consistency with our more conservative fiducial results. 

\subsubsection{High-mass, high-spin population}
\label{sec:high_spin_high_mass}
The transition into the higher mass regime begins at $m_1 = \JSONParseValue{\FiducialMacros}{mass_breakpoints[2]} \, M_\odot$, which we inferred previously with a two-component population model using GWTC-4 data~\citep{Hussain:2026pfm}.
This is consistent with the transition mass identified by~\cite{Pierra:2024fbl,Tong:2025wpz,Antonini:2025ilj,Banagiri:2025dmy,Sridhar:2025kvi,LIGOScientific:2026ctl}, and attributed to the possible onset of (pulsational) pair instability~\citep{Fowler:1964zz,Barkat:1967zz,Woosley:2016hmi,Farmer:2019jed,Woosley:2021xba}.
Such processes have been predicted to prevent \ac{BH} formation over a range of masses, resulting in a gap.
The evidence for a sharp pair-instability gap and its precise location nevertheless remain sensitive to the population model~\citep{Ray:2025xti}.

Given the high values of $m_1$ potentially inside the pair instability mass gap, high spins, and wide range of mass ratios, a compelling formation channel for this population is via hierarchical mergers~\citep{Gerosa:2017kvu,Rodriguez:2019huv,Gerosa:2021mno,Li:2026iae,Guttman:2026cnv}, where one or more of the \acp{BH} is the remnant of a previous merger, and hence would be rapidly spinning.
High-mass systems exhibit additional features suggestive of hierarchical formation, including larger spin magnitudes, more nearly isotropic spin orientations, and broader mass-ratio distributions~\citep{Antonini:2024het,Antonini:2025ilj,Antonini:2025zzw,Hussain:2026pfm,Banagiri:2025dmy,LIGOScientific:2026ctl,Plunkett:2026pxt}.

However, there is some tension between our results and naive expectations from a hierarchical merger scenario.
Given that the majority of \acp{BH} at lower masses are inferred to have small or negligible spin we expect that a first generation plus second generation merger (\OneGTwoG{}) would have a large $\chi_1$ and small $\chi_2$.
Moreover, merger remnants receive recoil kicks with velocities often comparable to or exceeding the escape speeds of their proposed host environments.
Standard cluster models therefore predict \OneGTwoG{} mergers to dominate over \TwoGTwoG{} mergers, unless mass segregation and preferential dynamical pairing substantially enhance the latter channel~\citep{Rodriguez:2016vmx,Rodriguez:2019huv,Kimball:2020opk,Zevin:2022bfa,Vijaykumar:2026zjy}.

Our inferred population displays a broad range of $\chi_2$ values, so if large spins arise from past mergers we favor a scenario where there is a similar fraction of second plus second generation (\TwoGTwoG{}) mergers to \OneGTwoG{} mergers.
If the spin tilt distribution of this population is not isotropic, then a hierarchical scenario would also require additional environmental effects to break isotropy, such as gas torques and repeated interactions in disks of active galactic nuclei~\citep{Yang2019,Tagawa2021,2024MNRAS.531.3479M,Li:2025iux}.

Explanations other than the hierarchical scenario are possible.
Strong tidal interactions, which spin up the cores of both stars in close binaries, could produce large spins on either \ac{BH}~\citep{Mandel:2015qlu,deMink:2016vkw,Marchant:2016wow,Riley:2020btf};
in that case, we would expect the spins to be correlated with each other and more nearly aligned with the angular momentum.
Other alternatives include accretion-driven growth and spin-up in \ac{AGN} disks~\citep{Yang2019,McKernan:2019beu,Tagawa2021,Vaccaro:2023cwr,Bartos:2025pkv,Bartos:2026xlt} or dense star clusters~\citep{Kiroglu:2025vqy}, stellar collisions in young clusters~\citep{DiCarlo:2019pmf,Kremer:2020wtp}, the collapse of rapidly rotating very massive stars~\citep{Uchida:2019gjx,Gottlieb:2025ugy,Shibata:2025lde}, uncertain stellar physics that shifts the onset of the pair-instability gap~\citep{Croon:2025gol}, and primordial \acp{BH}~\citep{DeLuca:2020sae,DeLuca:2025fln}.

\subsubsection{Low-mass, high-spin population}
\label{sec:high_spin_low_mass}

Finally we consider the low-mass, high-spin population.
Several recent analyses report a fourth \ac{BBH} subpopulation at $15 \lesssim m_1/M_\odot \lesssim 25$, characterized by unequal component masses and large primary spins suggestive of repeated mergers~\citep{Ray:2026fdm,Tong:2025xir,Flanagan:2026btu,Guttman:2026cnv,LIGOScientific:2026ctl}.
Our analysis assigns \ac{BBH} candidates with these characteristics, such as GW241011 and GW241110~\citep{LIGOScientific:2025brd},
to the same subpopulation as the high-mass, high-spin \acp{BBH}.
Thus, while our results support the presence of this feature, we do not find evidence that its mass-ratio and spin distributions differ from those of the high-mass systems.
As such, these may arise from a common channel as the high-mass \acp{BBH} active across a continuum of masses, such as hierarchical mergers~\cite[e.g.,][]{Rodriguez:2019huv,Li:2026iae}.
It is also conceivable that these binaries could arise from an entirely separate channel, which our model cannot yet separate from the high-mass population.

We find that the rapidly spinning subpopulation $C$ accounts for ${\sim} 10\%$ of \acp{BBH} between $15$ and $25\, M_\odot$ (Figure~\ref{fig:fraction}), which is broadly consistent with recent analyses which have quoted differing ranges of branching fractions depending on modeling choices.
Analyses that treat the $15$--$25\, M_\odot$ window with its own mass-ratio and spin distribution, independent of the high-mass systems~\citep{Ray:2026fdm,Guttman:2026cnv}, do not impose a belief that the $(q,\chi)$ structure should be shared between the two mass scales; our model shares that information across mass, so the fraction assigned to subpopulation $C$ near $20\, M_\odot$ is set jointly by both regimes.
In spite of modeling differences, \citet{Guttman:2026cnv} finds a mixture fraction for their low-mass, high-spin (labeled \texttt{horizontal}) component of 13\%, consistent with ours.

When assuming two possible $\chi_\mathrm{eff}$ distributions, one flat and broad and the other Gaussian, with mass-dependent mixing fractions, both \citet{LIGOScientific:2026ctl} and~\citet{Flanagan:2026ayy} find anywhere from $10\%$ to $100\%$ of \acp{BBH} between $10$ and $20\, M_\odot$ in the broad-$\chi_{\rm eff}$ class.
These approaches parametrize $\chi_{\rm eff}$ rather than the component spin magnitudes and tilts, and have no analogue of our third moderately spinning subpopulation, which can absorb some of the spinning, lower mass \acp{BBH} in our model.
The closest comparison our model offers is the total fraction of \acp{BBH} with non-negligible spin in this mass range, $1 - \eta_A(m_1=20M_\odot)$ for which we infer, e.g., ${\sim}10$--$40\%$ at $90\%$ credibility, consistent with the lower portion of the range in \cite{LIGOScientific:2026ctl}.

\subsection{Model Variations}
\label{sec:model_variations}
We have explored multiple variations to our model and find that our results are robust across these. 

Firstly, under our model assumptions, the data do not yet resolve another feature in $(q,\boldsymbol{\chi}_1, \boldsymbol{\chi}_2)$ that is sufficiently distinct from the three already inferred.
We allow for a fourth subpopulation (D) with its own mixing fraction $\eta_D(m_1)$ and $(q,\boldsymbol{\chi}_1, \boldsymbol{\chi}_2)$ distribution, allowing for three transitions in the $\eta_k(m_1)$.
If the resultant mass-ratio and spin distribution of the fourth population are found to be a weighted mixture of the other three, it is simply degenerate with a three-subpopulation result.
Indeed, we find that the newly added subpopulation simply blends the spin distributions of the medium- and high-spin populations while following the same mass fraction $\eta(m_1)$ as our fiducial high-spin population.

Secondly, we find that the data do not require more than three transition regions to describe $\eta(m_1)$ for three subpopulations.
Specifically motivated by the second peak in $\eta_C(m_1)$ (Figure~\ref{fig:fraction}) at lower masses, we explored allowing for four flat regions with three transitions in the three mass fractions $\eta_k(m_1)$.
This allows for the greater flexibility for a given population to be concentrated in multiple regions in $m_1$ space, since the number of peaks and troughs in $\eta(m_1)$ are directly dictated by the number of flat regions.
Overall the effects of this change in our model are small.
Our results remain qualitatively unchanged, although we find that the $\ac{PPD}$ of the fraction of the high-spin population in the range 
$m_1 \sim 15$--$25\, M_\odot$
broadens, allowing for somewhat higher fractions (see Appendix~\ref{appx:FurtherResults});
the spin magnitude and cosine tilt \acp{PPD} remain the same. 
With the additional flexibility the mass-fraction \acp{PPD} of all subpopulations widen at large primary masses $m_1 \gtrsim 100 M_\odot$, where only GW231123\_135430 and GW231028\_153006 \citep{LIGOScientific:2025rsn,LIGOScientific:2026wfs} are informative  (essentially reverting towards the prior at high masses).

Lastly, allowing for off-diagonal components in $\boldsymbol{\Sigma}_\chi^k$ yields nearly identical \acp{PPD}, but makes sampling more difficult.
We attribute the sampling challenges to degeneracies that arise in fitting the highly concentrated low-spin population with a truncated normal if correlations are allowed.
Even restricting to a diagonal $\boldsymbol{\Sigma}_\chi^k$, we still recover correlated structure in the spin-magnitude \acp{PPD}, as evidenced by subpopulation $B$ in Figure~\ref{fig:SpinPPDs}; this arises due to narrow features whose locations in $(\chi_1,\chi_2)$ are anti-correlated at the level of the hyperposterior, as can be gleaned in Figure~\ref{fig:medium_spin_hyperparams}.
Additionally, when analyzing the dataset with a reduced spin tilt model using only a truncated normal for each subpopulation, we find essentially the same results as illustrated in Figure~\ref{fig:SpinPPDs}.

As discussed in Section~\ref{sec:high_spin}, we have also explored widening the hyperpriors controlling the mass ratio distributions in our fiducial model.
This causes the mass-ratio distribution of the high-spin population to shift somewhat towards lower mass ratios, introduces greater uncertainty in the \acp{PPD} of the spin parameters, and increases the high-spin fraction at moderate masses $15\, M_\odot\lesssim m_1 \lesssim 40 \, M_\odot$, as discussed in more detail in Appendix~\ref{appx:FurtherResults}.
This also increases the uncertainty in our likelihood estimates, but overall does not change our conclusions.

\section{Conclusions}
\label{sec:Conclusions}

In this work, we analyzed the 259 \acp{BBH} in GWTC-5~\citep{LIGOScientific:2026ctl}, seeking distinct subpopulations of \acp{BBH} in the latest \ac{LVK} catalog~\citep{LIGOScientific:2026wfs}.
Using a flexible model which allows for multiple, overlapping subcomponents in primary mass space with distinct spin and mass ratio distributions, we find strong evidence for at least three subpopulations.
These neatly separate in the two-dimensional space of component spin magnitudes into populations with low, medium, and high spins.
If we allow for a fourth subpopulation, it does not acquire a distinct mass-ratio and spin distribution, but instead becomes degenerate with the components already present.
One distinct aspect of our approach is the use of methods to control the variance in estimates of the population likelihood, allowing for the assessment of highly concentrated populations~\citep{Hussain2025,Hussain:2025llf}.
Another is to allow the component spins to be drawn from independent distributions, with correlations among some parameters. 
This allows for a greater range of population models than those based on effective spin quantities, and models based on individual component spins are easily compared to astrophysical predictions.

The low-spin population dominates at low masses and makes up most of the intermediate mass \acp{BBH} around ${\sim} 35 \, M_\odot$, and our results show strong evidence that these binaries have negligible spin. 
This is compatible with a population where efficient angular momentum transport during stellar evolution leads to small natal spins.
They may be dynamically assembled or arise as isolated field binaries which avoid binary interactions that spin up either \ac{BH}.
The medium-spin component is present primarily at low masses, with preferentially aligned spins and modest spin magnitudes $\chi_i \sim 0.2$.
These appear to be naturally explained by isolated binary evolution in cases where mass transfer or tidal spin up results in \acp{BH} with modest natal spins.
The highly spinning population makes up nearly all of the \acp{BBH} above ${\sim} 60 M_\odot$ and also provides a subdominant contribution peaking at ${\sim} 20 \, M_\odot$ and potentially extending through the intermediate mass region.
It carries some of the expected imprints of hierarchically formed binaries: large spins on the primary,
$\chi_1 = \JSONParseValue{\FiducialMacros}{subpop_ppds.c.chi_1}$, a nearly flat mass-ratio distribution supporting unequal mass binaries, and a distribution of spin tilts consistent with isotropic spins.
Other features of the high-spin population, such as a wide range of spins on the secondary, $\chi_2 = \JSONParseValue{\FiducialMacros}{subpop_ppds.c.chi_2}$ and hints of structure in the spin tilts, point towards alternate or more complex formation scenarios.

The principal result of our component-spin analysis is that the population structure is sharper than is apparent from the number of inferred subpopulations alone.
The binaries dominating the population near $35\,M_\odot$ are consistent with both component black holes having negligible spin, while a distinct population with moderate, preferentially aligned spins overlaps them at lower masses.
At high masses, the spin signature is instead carried most clearly by a rapidly spinning primary, with a much broader range of secondary spins.

Looking ahead, as data from the remainder of the fourth \ac{LVK} observing run is analyzed and future observing runs are carried out, we expect our \ac{GW} catalogs to continue to rapidly grow, increasing our ability to resolve population features.
Of special interest is the question of whether with a greater number of detections the high-spin population resolves into further subpopulations. For example, more data could lead to a differentiation of the lower mass and higher mass components of the high spin population, or a further division into populations that carry separate imprints of \OneGTwoG{} and \TwoGTwoG{} mergers, or a division into dynamically formed, isotropic versus chemically homogeneous, relatively aligned binaries.
Regardless, the increasing number of \acp{BBH} detections in the future will surely lead to a greater understanding of the lives and deaths of massive stars, and the still uncertain astrophysical processes that govern the formation and characteristics of \acp{BH}.
\\

\noindent 
We thank Aditya Vijaykumar, Soumendra Kishore Roy, Barry McKernan, Noah Wolfe, Jack Heinzel, Salvatore Vitale, Konstantin Leyde, Saavik Ford and Will Farr for helpful discussions.
This work carries preprint numbers LIGO-P2600556 and UT-WI-27-2026.
AZ was supported by NSF Grant PHY-2308833.
The Flatiron Institute is a division of the Simons Foundation.
This material is based upon work supported by NSF's LIGO Laboratory which is a major facility fully funded by the National Science Foundation.
The authors are grateful for computational resources provided by the LIGO Laboratory and supported by NSF Grants PHY-0757058 and PHY-0823459.
This research has made use of data or software obtained from the Gravitational Wave Open Science Center (gwosc.org), a service of the LIGO Scientific Collaboration, the Virgo Collaboration, and KAGRA. 

\textit{Software: } This work was implemented using \textsc{python }\citep{Python3} and  \textsc{Julia} \citep{Julia-2017}. 
We made use of the following python packages: \textsc{numpy }\citep{numpy},  \textsc{scipy }\citep{scipy},  \textsc{pandas }\citep{pandas},  \textsc{matplotlib}\citep{matplotlib} \textsc{numpyro}.
Similarly, this work utilized the following julia packages: \textsc{QuadGK.jl} \citep{quadgk}, \textsc{Distributions.jl} \citep{JSSv098i16,Distributions.jl-2019}, \textsc{DataFrames.jl}~\citep{JSSv107i04} and \textsc{Roots.jl}~\citep{Roots.jl}.

\appendix

\section{Details on methods and population models}
\label{appx:Methods}

\subsection{Population model and hyperpriors}
\label{appx:model}

Here we give further details on the population model introduced in Section~\ref{sec:Methods}, and list our priors on all hyperparameters in Table~\ref{tab:hyperpriors}.
Recalling Eq.~\eqref{eq:fullmodel}, the full model factorizes as
\begin{align}
    p(\boldsymbol{\theta} \mid \boldsymbol{\Lambda})  = &
    p(z\mid\boldsymbol{\Lambda}_z)
    p(m_1\mid\boldsymbol{\Lambda}_m)
    \sum_{k \in \{A,B,C\}}
    \eta_k(m_1\mid\boldsymbol{\Lambda}_{\eta}^k)\, p(q,\boldsymbol{\chi}_1,\boldsymbol{\chi}_2\mid\boldsymbol{\Lambda}_\chi^k)\,.
    \label{eq:fullmodel_appx}
\end{align}
For the shared primary-mass distribution $p(m_1\mid\boldsymbol{\Lambda}_m)$ and redshift distribution $p(z\mid\boldsymbol{\Lambda}_z)$ we adopt the standard fiducial Strongly Modeled model used in the GWTC-5 population analysis~\cite{LIGOScientific:2025pvj,LIGOScientific:2026ctl}: a broken power law plus two Gaussian peaks in $m_1$, and a power-law-in-$(1+z)$ merger-rate evolution.

Each subpopulation $k \in \{A,B,C\}$ is assigned a weight $\eta_k(m_1)$ that is piecewise-constant across three flat regimes of primary mass, connected by two linear transitions.
The regimes and transitions are delimited by four break masses $10 \leq m^\eta_1 \leq m^\eta_2 \leq m^\eta_3 \leq m^\eta_4$, shared across all three subpopulations, and each subpopulation takes the constant value $\eta_k^{(r)}$ in flat regime $r \in \{0,1,2\}$:
\begin{equation}
\label{eq:eta_model}
\eta_k(m_1) =
\begin{cases}
\eta_k^{(0)}, & m_1 < m^\eta_1, \\[2pt]
\mathrm{lerp}\big(\eta_k^{(0)},\eta_k^{(1)}\big), & m^\eta_1 \leq m_1 < m^\eta_2, \\[2pt]
\eta_k^{(1)}, & m^\eta_2 \leq m_1 < m^\eta_3, \\[2pt]
\mathrm{lerp}\big(\eta_k^{(1)},\eta_k^{(2)}\big), & m^\eta_3 \leq m_1 < m^\eta_4, \\[2pt]
\eta_k^{(2)}, & ~~~~~~~~m_1 \geq m^\eta_4, \\[2pt]
\end{cases}
\end{equation}
where $\mathrm{lerp}\big(\eta_k^{(r)},\eta_k^{(r+1)}\big)$ denotes linear interpolation in $m_1$ between the two flat levels.
In each regime $r$ the three levels are properly normalized, $\sum_{k} \eta_k^{(r)} = 1$, and since Eq.~\eqref{eq:eta_model} performs simple linear interpolation, the constraint $\sum_{k}\eta_k(m_1) = 1$ holds at every $m_1$.
As a result of our priors and constraints, at fixed $m_1$ the prior distribution on the $\eta_k(m_1)$ is uniform within the constraint surface (drawn from a flat Dirichlet distribution).

Within each subpopulation the mass ratio and spin-magnitudes are modeled as truncated normals, whereas the spin-tilts are modeled as a mixture of a bi-variate truncated normal and a uniform distribution,
\begin{align}
p(q,\boldsymbol{\chi}_1,\boldsymbol{\chi}_2\mid\boldsymbol{\Lambda}_\chi^k) = &\
    \mathcal{N}_{[0,1]}(q\mid\mu_q^k,\sigma_q^k)\,
\mathcal{N}_{[\boldsymbol{0},\boldsymbol{1}]}(\chi_1,\chi_2\mid\boldsymbol{\mu}_\chi^k,\boldsymbol{\Sigma}_\chi^k) \\ \notag &\left[(1-\zeta_{\cos \theta}^k)U_{[-1,1]}(\cos\theta_1)\, U_{[-1,1]}(\cos\theta_2) + \zeta_{\cos\theta}^k \mathcal{N}_{[-\boldsymbol{1},\boldsymbol{1}]}(\cos \theta_1,\cos \theta_2\mid\boldsymbol{\mu}_\theta^k,\boldsymbol{\Sigma}_\theta^k)\right]\,,
\end{align}
where the subscripts denote truncation of the domain: $q$ to $[0,1]$, the spin magnitudes to the unit square $[0,1]\times[0,1]$, and the cosine tilts to $[-1,1]\times[-1,1]$.
The spin-tilt covariance matrix takes the form
\begin{equation}
	\bm{\Sigma}  = \begin{pmatrix}
		\sigma_1^2 & \rho\, \sigma_1 \sigma_2 \\
		\rho\, \sigma_1 \sigma_2 &  \sigma_2^2
	\end{pmatrix},
\end{equation}
so that each subpopulation supports an independent correlation between the two cosine tilts ($\rho_{\cos\theta}^k$).
For the spin magnitudes we assume a diagonal covariance ($\rho = 0$ in the above matrix) to ease the sampling.
Note that due to truncation, the correlation parameter $\rho$ in $\bm{\Sigma}$ is not the true correlation of the bivariate truncated Gaussian, i.e.~$\rho \neq\hat{\rho} = \text{Cov}[\chi_1, \chi_2]/(\text{Var}[\chi_1] \text{Var}[\chi_2])^{1/2}$.

We adopt broad uniform priors on the mass-ratio, spin-magnitude, and spin-tilt parameters of every subpopulation.
The spin-magnitude correlations are fixed to zero, while the correlations between the two cosine tilts vary freely over their stated prior range.
The four break masses have a joint flat prior over the ordered region
$10\,M_\odot\leq m^\eta_1\leq m^\eta_2\leq m^\eta_3\leq m^\eta_4\leq200\,M_\odot$, and the per-regime fractions have flat Dirichlet priors on the simplex.
The mixing-fraction, mass-ratio, and spin priors are collected in Table~\ref{tab:hyperpriors}.

\begin{figure}[tb]
    \centering
    \includegraphics[width=\columnwidth]
    {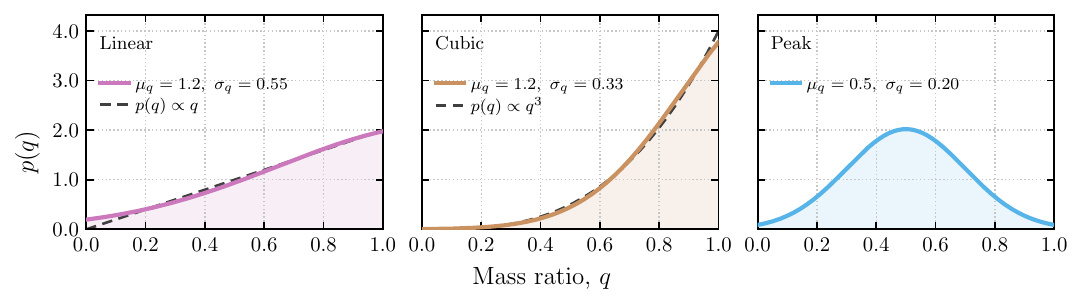}
    \caption{
    Examples of distributions accessible with our truncated-normal
    mass-ratio model.
    When $\mu_q>1$, the portion of the Gaussian retained within
    $0<q\leq1$ can approximate a power law.
    The choices $(\mu_q,\sigma_q)=(1.2,0.55)$ and $(1.2,0.33)$
    produce distributions similar to $p(q)\propto q$ and
    $p(q)\propto q^3$, respectively.
    When $\mu_q$ lies within $[0,1]$, the model can instead
    describe an interior Gaussian-like peak, as illustrated by
    $(\mu_q,\sigma_q)=(0.5,0.20)$.
    }
    \label{fig:mass-ratio-model-flexibility}
\end{figure}
Our truncated-normal model for the mass ratio encompasses power-law-like, peaked, and nearly uniform distributions within a single functional family, as illustrated in Figure~\ref{fig:mass-ratio-model-flexibility}.
The parameters $\mu_q^k$ and $\sigma_q^k$ specify the location and width of the underlying untruncated Gaussian, rather than the moments of the distribution after truncation to $q\in[0,1]$.
Allowing $\mu_q^k>1$ in our hyperprior therefore exploits the flexibility introduced by truncation: when the Gaussian mean lies above $[0,1]$, the retained tail can closely approximate a power law, $p(q)\propto q^\beta$, over a broad range of slopes, with $\beta\simeq(\mu_q^k-1)/(\sigma_q^k)^2$.
For $\mu_q^k<1$, the distribution can instead develop an interior peak, while sufficiently large $\sigma_q^k$ produces an approximately uniform distribution.

We impose no ordering constraints on the component labels during sampling, so the raw labels are exchangeable and may differ among runs.
In post-processing, for each posterior sample we define the score
$s_k \equiv \mu_{\chi_1}^k+\mu_{\chi_2}^k$
and permute all subpopulation-specific parameters so that
$s_A < s_B < s_C$.
Within individual chains the required permutation is typically constant across samples, with $\lesssim 0.1\%$ of samples swapping labels within a chain for our fiducial model.
This swapping occurs naturally when there is posterior support in the region where two subpopulations are almost identical based on the label defining score $s_k$.
This indicates that the data resolve three distinct features in $(q,\boldsymbol{\chi}_1,\boldsymbol{\chi}_2)$ with such confidence that there is very little posterior support in the region where any two distributions may look identical.

\begin{table*}[t]
\centering
\begin{minipage}{0.48\textwidth}
\centering
\begin{tabular}{ll}
\hline
Parameter & Prior \\
\hline
\multicolumn{2}{l}{Mixing fractions $\eta_k(m_1)$} \\
$m^\eta_1,\dots,m^\eta_4$
& $U(10,200)$  \\
& $m^\eta_1 \leq m^\eta_2
\leq m^\eta_3 \leq m^\eta_4$ \\
$(\eta_A^{(0)},\eta_B^{(0)},\eta_C^{(0)})$
& $\mathrm{Dir}(1,1,1)$ \\
$(\eta_A^{(1)},\eta_B^{(1)},\eta_C^{(1)})$
& $\mathrm{Dir}(1,1,1),\;$ \\
$(\eta_A^{(2)},\eta_B^{(2)},\eta_C^{(2)})$
& $\mathrm{Dir}(1,1,1),\;$ \\
\hline
\multicolumn{2}{l}{Mass ratio \;(each $k \in \{A,B,C\}$)} \\
$\mu_q^k$ & $U(0.4,1.2)$ \\
$\sigma_q^k$ & $U(0.2,0.8)$ \\
\hline
\end{tabular}
\end{minipage}
\hfill
\begin{minipage}{0.48\textwidth}
\centering
\begin{tabular}{ll}
\hline
Parameter & Prior \\
\hline
\multicolumn{2}{l}{Spin magnitudes \;(each $k \in \{A,B,C\}$)} \\
$\mu_{\chi_1}^k,\;\mu_{\chi_2}^k$ & $U(0,1)$ \\
$\sigma_{\chi_1}^k,\;\sigma_{\chi_2}^k$ & $U(0,1)$ \\
\hline
\multicolumn{2}{l}{Spin tilts \;(each $k \in \{A,B,C\}$)} \\
$\mu_{\cos\theta_1}^k,\;\mu_{\cos\theta_2}^k$ & $U(-1,1)$ \\
$\sigma_{\cos\theta_1}^k,\;\sigma_{\cos\theta_2}^k$ & $U(0,2)$ \\
$\rho_{\cos\theta}^k$ & $U(-0.9,0.9)$ \\
$\zeta_{\cos\theta}^k$ & $U(0,1)$ \\
\hline
\end{tabular}
\end{minipage}
\caption{Hyperpriors for the mass-dependent mixing fractions and the subpopulation mass-ratio and spin distributions.
{\it Left:} the mass-dependent mixing fractions $\eta_k(m_1)$ (top) and the per-subpopulation mass ratio (bottom).
The four break masses share a single flat prior over the ordered region, and each regime's fractions are drawn from a flat Dirichlet so that $\sum_k \eta_k^{(r)} = 1$.
{\it Right:} the spin-magnitude and spin-tilt hyperparameters.
The mass-ratio and spin priors share the same form across the three subpopulations $k \in \{A,B,C\}$ and are listed once.
The primary-mass and redshift hyperpriors are given separately in Table~\ref{tab:massprior}.}
\label{tab:hyperpriors}
\end{table*}

\begin{table*}[t]
\centering
\begin{tabular}{lll}
\toprule
Parameter & Description & Prior \\
\midrule
\multicolumn{3}{l}{Primary-mass distribution} \\
$\alpha_1$ 
    & Power-law index below the break 
    & $U(-4,12)$ \\
$\alpha_2$ 
    & Power-law index above the break 
    & $U(-4,12)$ \\
$m_{\min}$ 
    & Minimum-mass scale 
    & $U(3,5)\,M_\odot$ \\
$\delta_m$ 
    & Width of the low-mass smoothing window 
    & $U(0,10)\,M_\odot$ \\
$\delta_{m,\mathrm{high}}$ 
    & Width of the high-mass smoothing window 
    & $U(0,20)\,M_\odot$ \\
$m_{\max}$ 
    & Maximum mass 
    & Fixed at $300\,M_\odot$ \\
$m_{\mathrm{break}}$ 
    & Break mass of the broken power law 
    & $U(20,50)\,M_\odot$ \\
$\mu_{m,1}$ 
    & Location of the lower-mass Gaussian peak 
    & $U(5,20)\,M_\odot$ \\
$\sigma_{m,1}$ 
    & Width of the lower-mass Gaussian peak 
    & $U(0.5,10)\,M_\odot$ \\
$\mu_{m,2}$ 
    & Location of the higher-mass Gaussian peak 
    & $U(25,60)\,M_\odot$ \\
$\sigma_{m,2}$ 
    & Width of the higher-mass Gaussian peak 
    & $U(1,10)\,M_\odot$ \\
$(\lambda_0,\lambda_1,\lambda_2)$
    & Weights of the broken power law and two Gaussian peaks
    & $\mathrm{Dir}(1,1,1)$ \\
\bottomrule
\end{tabular}
\caption{
Hyperpriors for the primary-mass and redshift distributions.
The primary-mass model is a mixture of a broken power law and two
Gaussian peaks, with mixture weights
$(\lambda_0,\lambda_1,\lambda_2)$.
The locations and widths of the Gaussian peaks are denoted by
$(\mu_{m,1},\sigma_{m,1})$ and
$(\mu_{m,2},\sigma_{m,2})$.
}
\label{tab:massprior}
\end{table*}

\section{Toy model injection and recovery}
\label{appx:Simulation}

\begin{figure*}[tb]
    \centering
    \includegraphics[width=1.0\columnwidth]{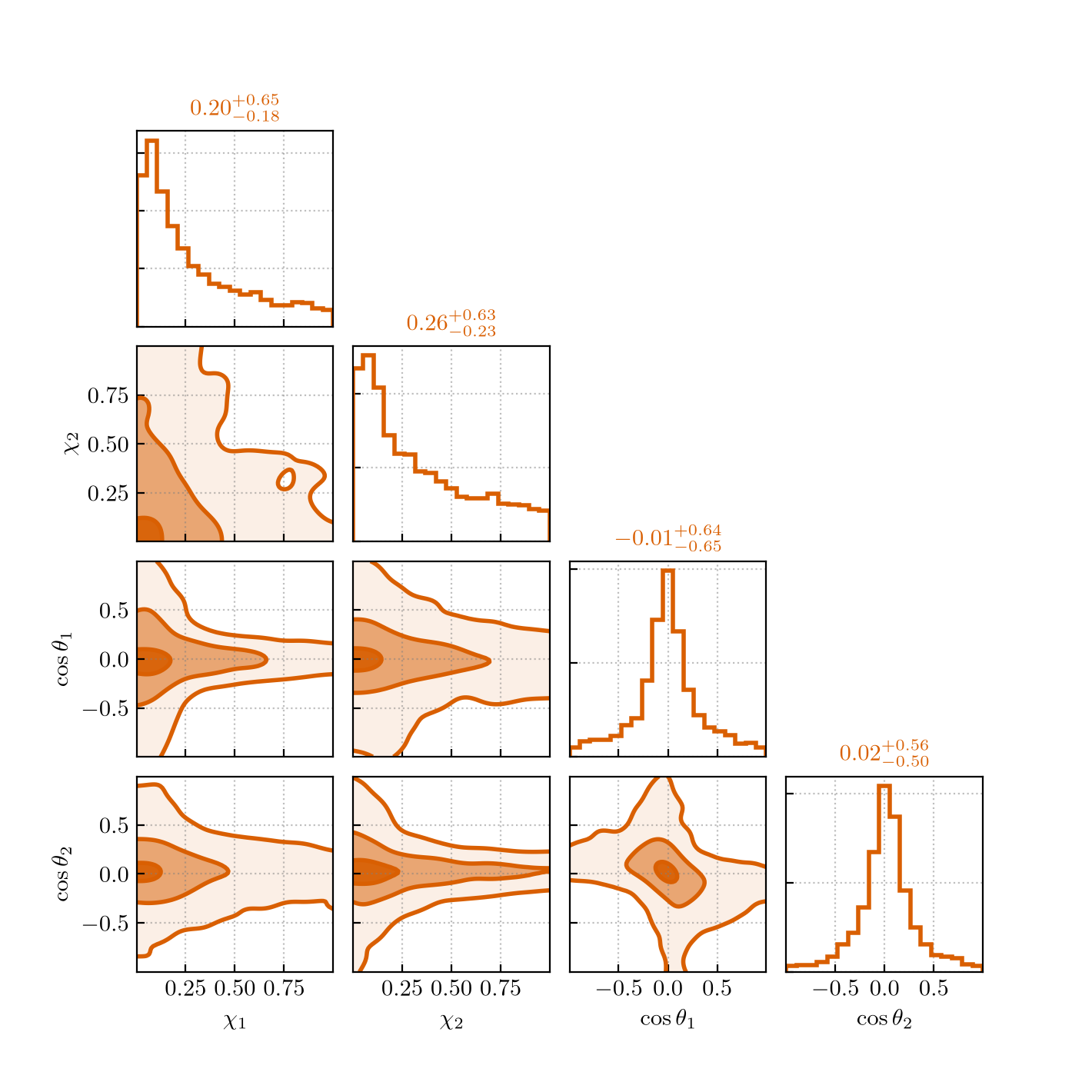}
    \caption{
    {Marginal \acp{PPD} for the spin magnitudes and cosine tilts for our toy injection and recovery of a nonspinning population. In this model we assume the \ac{BBH} spins are measured only through the effective spin $\chi_\mathrm{eff}$. The result is that while the population is nonspinning and so consistent with isotropic spin orientations, the spin tilts are highly peaked near the equatorial plane where there are no constraints but a large prior volume.} 
    }
    \label{fig:InjectionRecoveryToyModel}
\end{figure*}

For the low-spin population we find that the recovered tilt distributions are narrowly concentrated near the equatorial plane of the binary.
In this section we argue that this result remains consistent with a subpopulation with negligible spins, and we should not necessarily expect to recover isotropic spin tilts.
Utilizing a simple toy model, we show that inferring component spin population parameters ($\chi,\cos\theta$) for a negligibly spinning population when the detection process is mainly sensitive to only the effective combination $\chi_{\text{eff}}$, results in a consistent recovery of small spin magnitude paired with a recovery of a sharp peak in equatorial tilts.

Let $\boldsymbol{\chi} := (\chi_1,\chi_2,\cos\theta_1,\cos\theta_2)$ be a vector of the component spin variables and assume, for simplicity, equal masses $q=1$. 
We assume a detection process that is only sensitive to $\chi_{\text{eff}}$ with measurement uncertainty $\sigma^{\text{meas}} _{\chi_{\text{eff}}} = 0.1$, such that the PE likelihood is centered on the observed value $\chi_\mathrm{eff}^\mathrm{obs}$ and is simply
\begin{align}
    \mathcal{L}(d = \chi_{\text{eff}}^{\text{obs}}\mid \boldsymbol{\chi}_{1,2}) = \mathcal{N}(\chi_{\text{eff}}(\boldsymbol{\chi}) \mid \chi_{\text{eff}}^{\text{obs}},\sigma^{\text{meas}} _{\chi_{\text{eff}}})\,.
    \label{eq:simulation_pe_likelihood}
\end{align}
This mimics a measurement process only sensitive to $\chi_{\text{eff}}$. 

We simulate $N_{\text{event}}=150$ events drawn from a population with exactly zero spin magnitude and uniform cosine-tilts, and recover them with the population model
\begin{align}
    \label{eq:simulation_pop_model}
    \chi_i &\sim \mathcal{N}_{[0,1]}(\mu_{\chi_i}, \sigma_{\chi_i} ) \\
    \cos\theta_i  &\sim \mathcal{N}_{[-1,1]}(\mu_{\cos\theta_i}, \sigma_{\cos\theta_i})\,
\end{align}
with $i = 1,2$.
The hyperpriors on the above mimic those in Table~\ref{tab:hyperpriors}. 
The injected distribution is used to create $N_{\text{event}}=150$ values of $\chi_{\text{eff}}^{\text{true}} = 0$, and the measurement likelihood is used to generate a $\chi_{\text{eff}}^{\text{obs}}$ for each event.

To use \ac{HBI} to infer the properties of this negligibly spinning population, we require a technique to estimate the population likelihood for a given event $\mathcal{L}(d \mid \Lambda)$ accurately as $\sigma_{\chi_i} \to 0$. 
The process to achieve this is simple in this scenario, since we know the PE likelihood analytically for each given $\chi_{\text{eff}}^{\text{obs}}$.
We can draw samples from the population model, Equation~\eqref{eq:simulation_pop_model}, and use the PE likelihood, Equation~\eqref{eq:simulation_pe_likelihood}, as our weighting function to get a \ac{MC} estimate of $\mathcal{L}(d \mid \Lambda)$.
Specifically, we utilize the no-U-turn sampler \citep{2011arXiv1111.4246H} Hamiltonian \ac{MC} algorithm \citep{2011hmcm.book..113N,2017arXiv170102434B} as implemented in \textsc{numpyro}~\citep{phan2019composable,bingham2019pyro}, and maintain auto-differentiability of the described \ac{MC} estimator by transforming a fixed bank of uniform samples to the proposed population distribution using a differentiable map, and evaluating Equation~\eqref{eq:simulation_pe_likelihood}, which is differentiable. 
This way, we can probe narrow population features with no loss in the efficacy of our \ac{MC} estimator.

Figure~\ref{fig:InjectionRecoveryToyModel} displays the \ac{PPD} from the recovery of our $\chi_{\text{eff}} = 0$ injection.
The salient features from the recovery are a preference for small spin magnitudes, coupled with a sharp preference for equatorial tilts, as with our result for the low-spin population. 
This can be attributed to the fact that the spin components along the orbital angular momentum $\chi_{i,z} = \chi_i \cos \theta_i$, are typically better constrained than those that lie in the orbital plane, $\chi_{i,\perp} = \chi_i \sin \theta_i$.
In the case where the underlying spin is negligible, an event likelihood tends to rule out larger spin magnitudes for tilt angles closer to the angular momentum axis, and allows for larger spin magnitudes when the spins are tilted closer to the orbital plane.
Since spin distributions where the spin magnitudes have a larger range but tilts lying closer to the orbital plane have a larger prior volume in hyperparameter space, we can recover \acp{PPD} like that of the low-spin subpopulation in Figure~\ref{fig:SpinPPDs} even if the spins are isotropic, provided that their magnitudes are negligible.

For actual \ac{GW} event likelihoods, there are additional correlations among the spin parameters.
For example, the sum of aligned-spin components is better constrained than their differences, especially for binaries with nearly equal masses.
This accounts for the observed anti-correlation in the recovered spin-tilt distribution of the slowly spinning component in the right panel of Figure~\ref{fig:SpinPPDs}.
Further, large perpendicular components are constrained, if weakly, when precession effects are absent.
We attribute the lack of long tails seen in Figure~\ref{fig:InjectionRecoveryToyModel} towards high spin magnitudes when the spins are concentrated in the orbital plane to this.

\section{Further results}
\label{appx:FurtherResults}

Here we present additional results from both our fiducial model and variations on it.

\subsection{Fiducial model: three populations and three flat regions}

\begin{figure*}[tb]
    \centering
    \includegraphics[width = 0.9 \columnwidth]{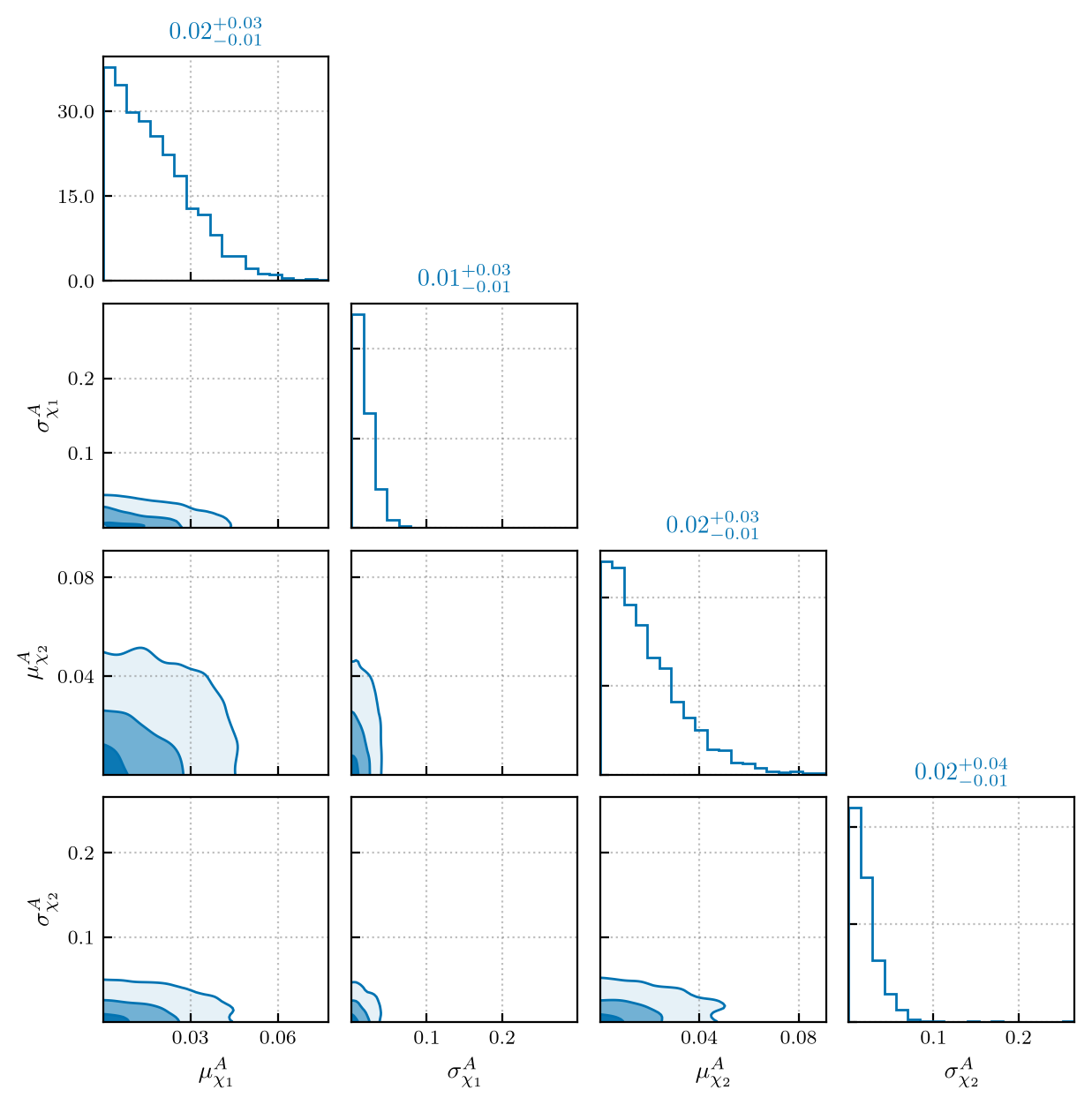}
    \caption{Hyperparameters for the spin distribution of the low-spin population.}
    \label{fig:low_spin_hyperparams}
\end{figure*}

\begin{figure*}[tb]
    \centering
    \includegraphics[width = 0.9 \columnwidth]{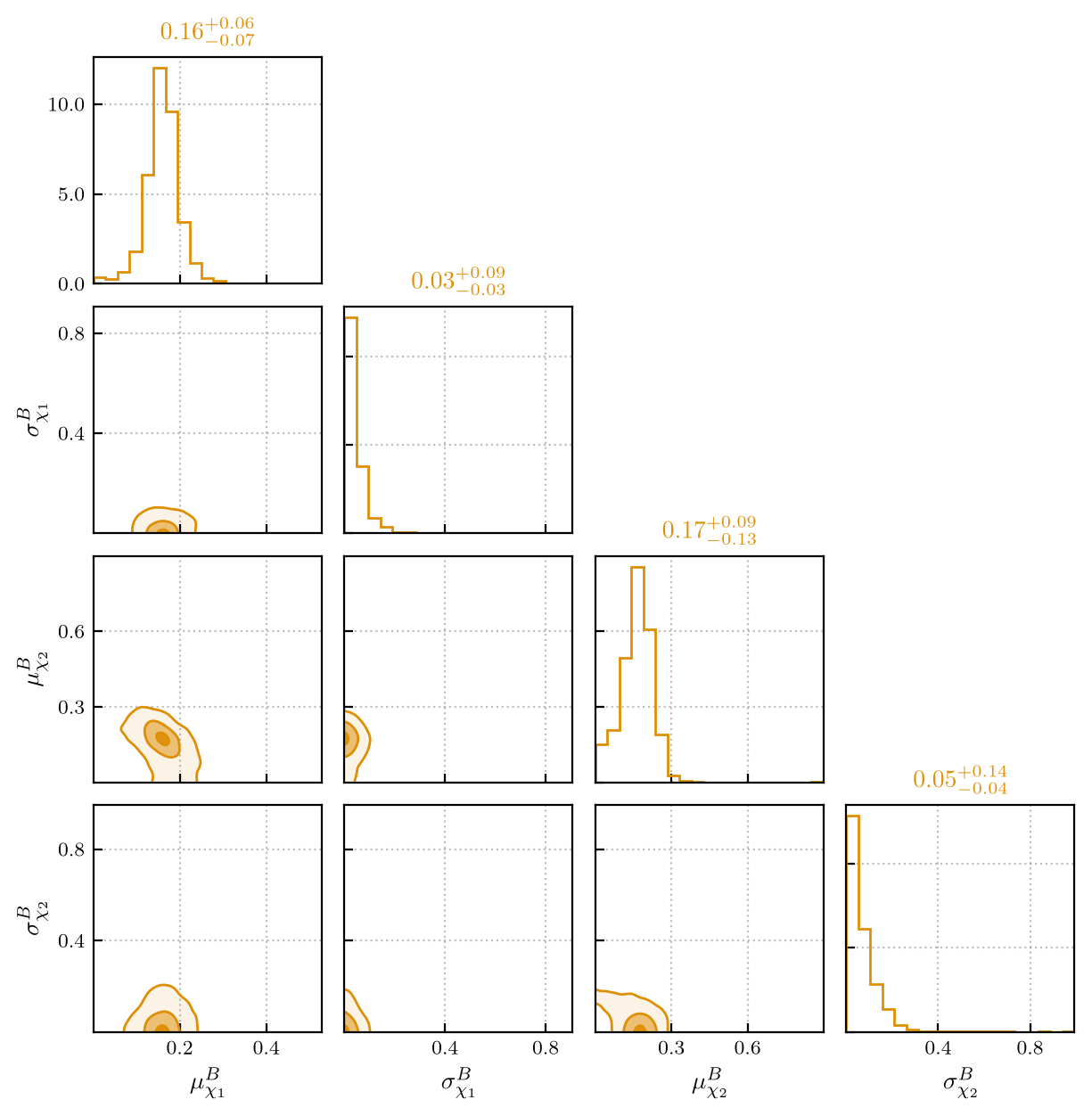}
    \caption{Hyperparameters for spin distribution of the medium-spin population.}
    \label{fig:medium_spin_hyperparams}
\end{figure*}

\begin{figure*}[tb]
    \centering
    \includegraphics[width = 0.9 \columnwidth]{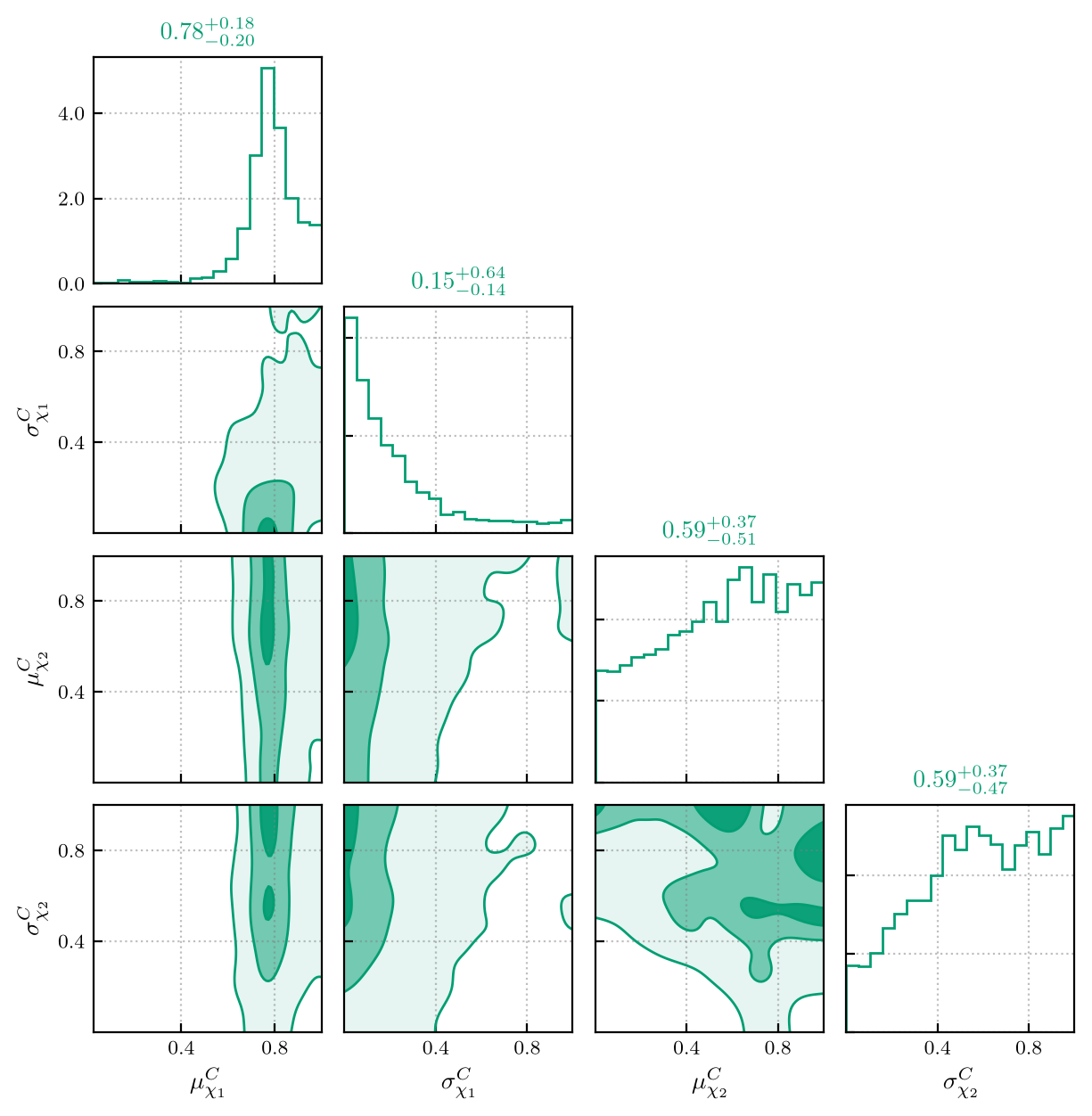}
    \caption{Hyperparameters for spin distribution of the high-spin population.}
    \label{fig:high_spin_hyperparams}
\end{figure*}

The hyperparameters controlling the three spin distributions are shown in Figures~\ref{fig:low_spin_hyperparams},~\ref{fig:medium_spin_hyperparams}, and~\ref{fig:high_spin_hyperparams}.
For the low-spin population (Figures~\ref{fig:low_spin_hyperparams}), one notable feature is that the point $\sigma^A_{\chi_1} = \sigma^A_{\chi_2} = 0$, corresponding to a population with negligible spread in both spin magnitudes, is consistent with the data and in fact preferred.
Such extremely narrow distributions are consistent with a range of location parameters $\mu^A_{\chi_1}$ and $\mu^A_{\chi_2}$ close to zero.
We strongly favor the hypothesis that all these \acp{BBH} have negligible spins, since $\log_{10} \mathcal B = \JSONParseValue{\FiducialMacros}{log_bayes_factors.in_favor_of_component_a_being_zero_spin}$ \citep{Jeffreys1939-JEFTOP-5,Kass01061995}.
Similarly, for the medium-spin population (Figures~\ref{fig:medium_spin_hyperparams}) a highly concentrated distribution of spin magnitudes is favored by the data, but with a spread in the posteriors of the location parameters.

More structure is visible in the hyperposteriors for the high-spin population, although these are broad.
Figure~\ref{fig:high_spin_hyperparams} shows that highly concentrated distributions of spin magnitudes are possible, but only when the location parameters of both spin magnitudes are both in the interval $0.65 \lesssim \mu^C_{\chi_1}, \mu^C_{\chi_2} \lesssim 0.85$,
At higher probabilities are cases where the secondary spin magnitude is broad, in which case a large range of $\mu^C_{\chi_2}$ values are consistent with the data.
Overall, the constraints on the secondary spin magnitude distribution are somewhat weak, and primarily rule out concentrated distributions at low spin values, as expected from the \ac{PPD} of Figure~\ref{fig:SpinPPDs}.

\begin{figure}[tb]
    \centering
    \includegraphics[width=0.9\columnwidth]{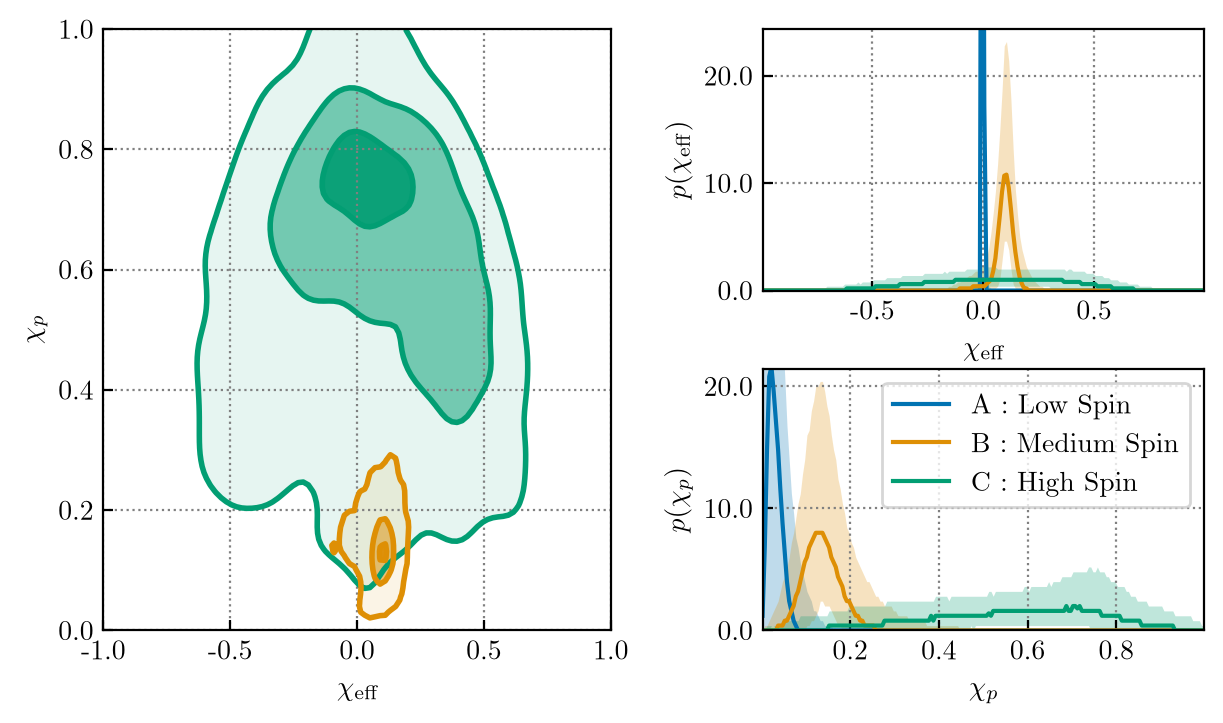}
    \caption{Inferred marginal \acp{PPD} for the effective spin parameters $\chi_\mathrm{eff}$ and $\chi_p$. The concentration of the $\chi_\mathrm{eff}$ distribution for the low-spin population is enhanced by a singular Jacobian in the transformation from spin magnitudes and tilts to the effective spin parameters. Due to the very narrow extent of the posteriors of the low-spin population, which is not well resolved by 2D kernel density estimation, we display it only for the 1D marginals.}
    \label{fig:effective_spin_PPDs}
\end{figure}

In this work we have focused on the spin magnitudes and tilt angles, which our model is built on.
From these and our mass inferences, we derive the inferred distribution of effective spin $\chi_\mathrm{eff}$ and effective precessing spin $\chi_p$.
These are given in Figure~\ref{fig:effective_spin_PPDs}.
We note that the inferred $\chi_\mathrm{eff}$ distribution for the low-spin population is highly concentrated to small values due to a combination of our measurement of the spin magnitudes and tilts and a strong Jacobian factor when mapping to the effective spin~\citep{Callister:2021gxf}.
The two-dimensional marginal \ac{PPD} for the high-spin population displays hints of the expected crescent shape suggested by~\cite{Plunkett:2026pxt}, which may be expected from hierarchical mergers.

\begin{figure}[tb]
    \centering
    \includegraphics[width=0.9\textwidth]{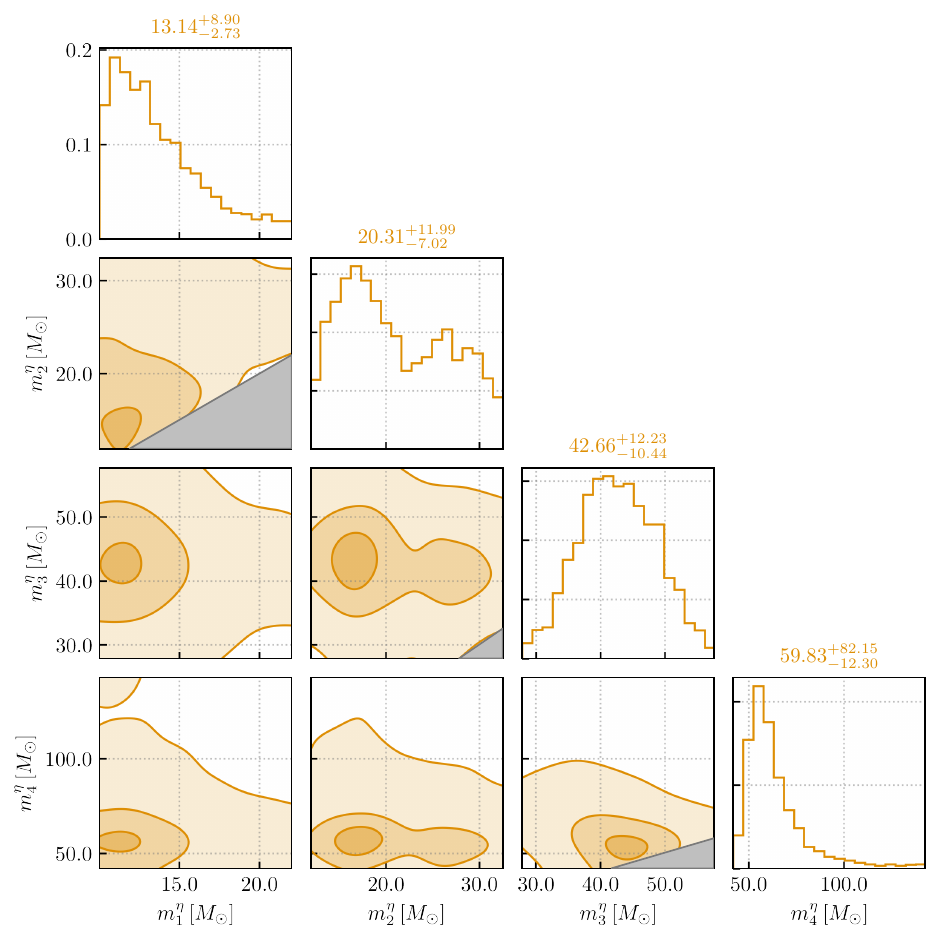}
    \caption{Inferences for the mass breakpoints $m^\eta_j$ where the transitions between flat fractions begin and end.
    The point $m^\eta_3$ marks the transition towards the high mass regime where the high spin subpopulation strongly dominates, with the transition ending at $m^\eta_4$.}
    \label{fig:mass_break_hyperparameters}
\end{figure}

The mass breakpoints where the fractions $\eta_k(m_1)$ transition between flat distributions likely carry the imprint of astrophysical and environmental properties.
Figure~\ref{fig:mass_break_hyperparameters} gives our posteriors over these hyperparameters.
Of especial interest are the breakpoints marking the transition to the high-mass region where the high-spin population dominates.
The transition starts at $m^\eta_3 = \JSONParseValue{\FiducialMacros}{mass_breakpoints[2]} \, M_\odot$, consistent with the mass scale attributed by previous analyses to the onset of pair instability processes~\citep[e.g.][]{Antonini:2025zzw,Tong:2025wpz}.
It ends at $m^\eta_4 = \JSONParseValue{\FiducialMacros}{mass_breakpoints[3]} \, M_\odot$, beyond which the high spin population dominates.

\begin{figure*}[t]
    \centering
    \includegraphics[width=\textwidth]{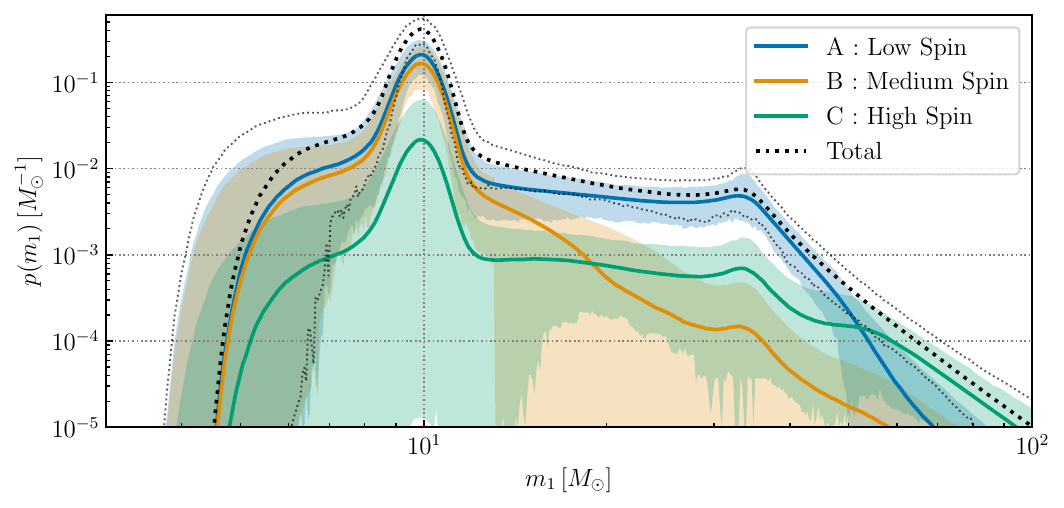}
    \caption{Inferred primary mass distribution for each of the three subpopulations, along with the total distribution. The mixing fractions $\eta_k(m_1)$ serve to shape and partially window the underlying common primary mass model, a broken power law plus two Gaussian peaks.
    }
    \label{fig:mass_ppd_by_component}
\end{figure*}

The inferred fractions $\eta_k(m_1)$ only give partial information about the distribution of \acp{BBH} across primary mass.
Figure~\ref{fig:mass_ppd_by_component} shows the inferred distribution of primary masses for each of the three subpopulations, and their total.
Since the global maximum of the distribution is the large peak at $\sim 10\, M_\odot$, where the low- and medium-spin populations have similar contributions, the absolute number of \acp{BBH} in the medium spin-population is larger.
We see that the steep decline of $\eta^B$ towards higher masses reduces the contribution of the medium-spin population below that of the high-spin population by around $20\,M_\odot$.
Correspondingly, the distribution of the low- and high-spin populations are flatter between the low-mass peak at $\sim 10\, M_\odot$ and the higher mass peak and power law break than the total distribution.
Beyond these features at $\sim 35 M_\odot$, the decline of $\eta^A$ steepens the fall off of the low-spin population, which falls below the high-spin near $\sim 60\, M_\odot$.
Beyond that the high-spin population dominates the population and follows the steeper, high-mass power law trend.

Finally, we can assess the probability for each individual binary system to lie in a given subpopulation.
Since these assignment charts are lengthy, we present them below in a dedicated sub-appendix, Appendix~\ref{sec:assignments}.
We also provide population-reweighted posteriors for the events using our fiducial model in our data release ~\citep{Hussain:2026data}.

\subsection{Three transitions and four flat regions}

\begin{figure*}[tb]
    \centering
    \includegraphics[width=0.9\textwidth]{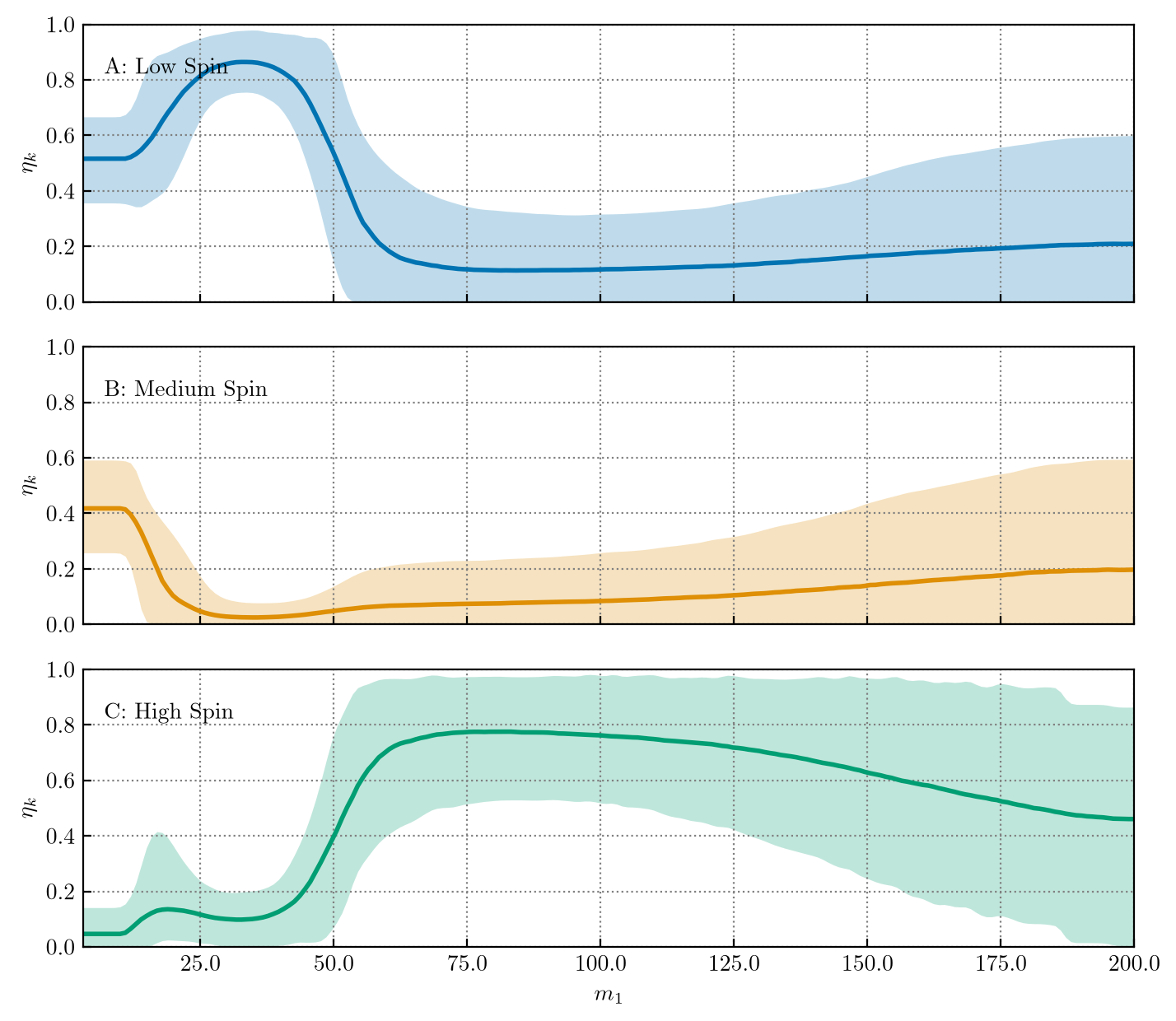}
    \caption{Inferred fractions $\eta_k(m_1)$ when allowing for four flat regions and three transitions between them. The primary difference from our fiducial result is a broadening of the \ac{PPD} of the high-spin populations near $m_1 \sim 20\,M_\odot$ and reversion towards the priors at the highest masses.}
    \label{fig:four_transitions_fractions}
\end{figure*}

As a modification to our fiducial model, we consider the effect of allowing for another flat region and corresponding transition, which can provide further substructure in the fractions $\eta_k$.
Figure~\ref{fig:four_transitions_fractions} shows our results. 
We find our results nearly unchanged, and the extra transition tends to occur at the highest masses $m_1 \gtrsim 100\,M_\odot$ where there are few observations.
At these masses the fractions broaden and revert towards the prior.
The fraction of high-spin \acp{BBH} at lower masses also broadens somewhat, with $\eta_C(20 M_\odot) = \JSONParseValue{\FourRegimeMacros}{fraction_at_masses.fraction_at_mass_20.C}$ as compared to the fiducial $\eta_C(20 M_\odot) = \JSONParseValue{\FiducialMacros}{fraction_at_masses.fraction_at_mass_20.C}$

\begin{figure*}[tb]
    \centering
    \includegraphics[width=0.48\columnwidth]{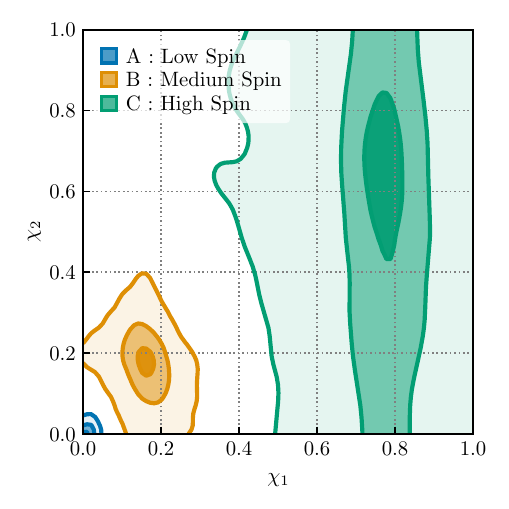}
    \includegraphics[width=0.48\columnwidth]{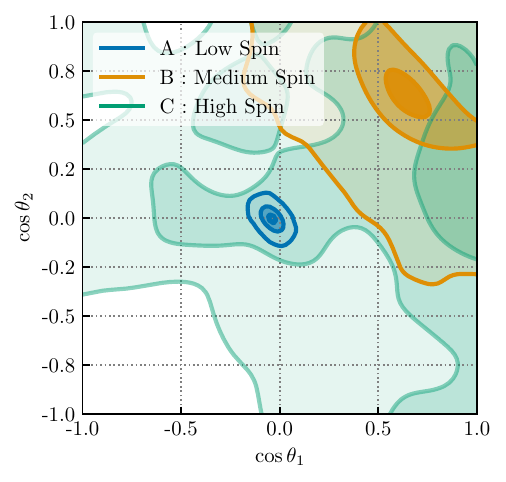}
    \caption{Inferred spin magnitude and cosine tilt distributions when allowing for four flat regimes in the mixing fractions $\eta_k(m_1)$. The results are essentially unchanged with the additional flexibility.
    }
    \label{fig:four_transitions_spins}
\end{figure*}

Meanwhile, our spin inferences remain essentially the same, as illustrated in Figure~\ref{fig:four_transitions_spins}.

\subsection{Four subpopulations}

\begin{figure*}
    \centering
    \includegraphics[width=0.47\columnwidth]{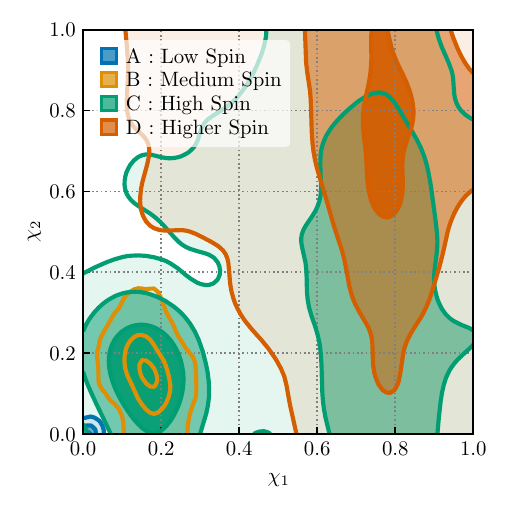}
    \includegraphics[width=0.48\columnwidth]{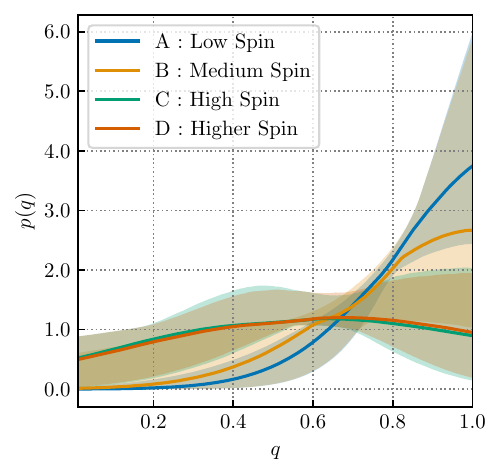}
    \caption{Inferred \acp{PPD} for the spin magnitude and mass-ratio distributions when allowing for a fourth subpopulation with mixing fraction $\eta^D(m_1)$. With the addition, two of the components are confused. {\it Left}: One population (high spin) supports spin magnitudes across two modes which correspond to those found by the medium- and high-spin population in our fiducial analysis, while the other (higher spin) broadly occupies the region of the fiducial high-spin population.
    {\it Right}: The distribution of mass ratio of the two confused components mirror each other and are similar to that if the fiducial high-spin population.}
    \label{fig:four_components_spins}
\end{figure*}

\begin{figure*}
    \centering
    \includegraphics[width=0.8\columnwidth]{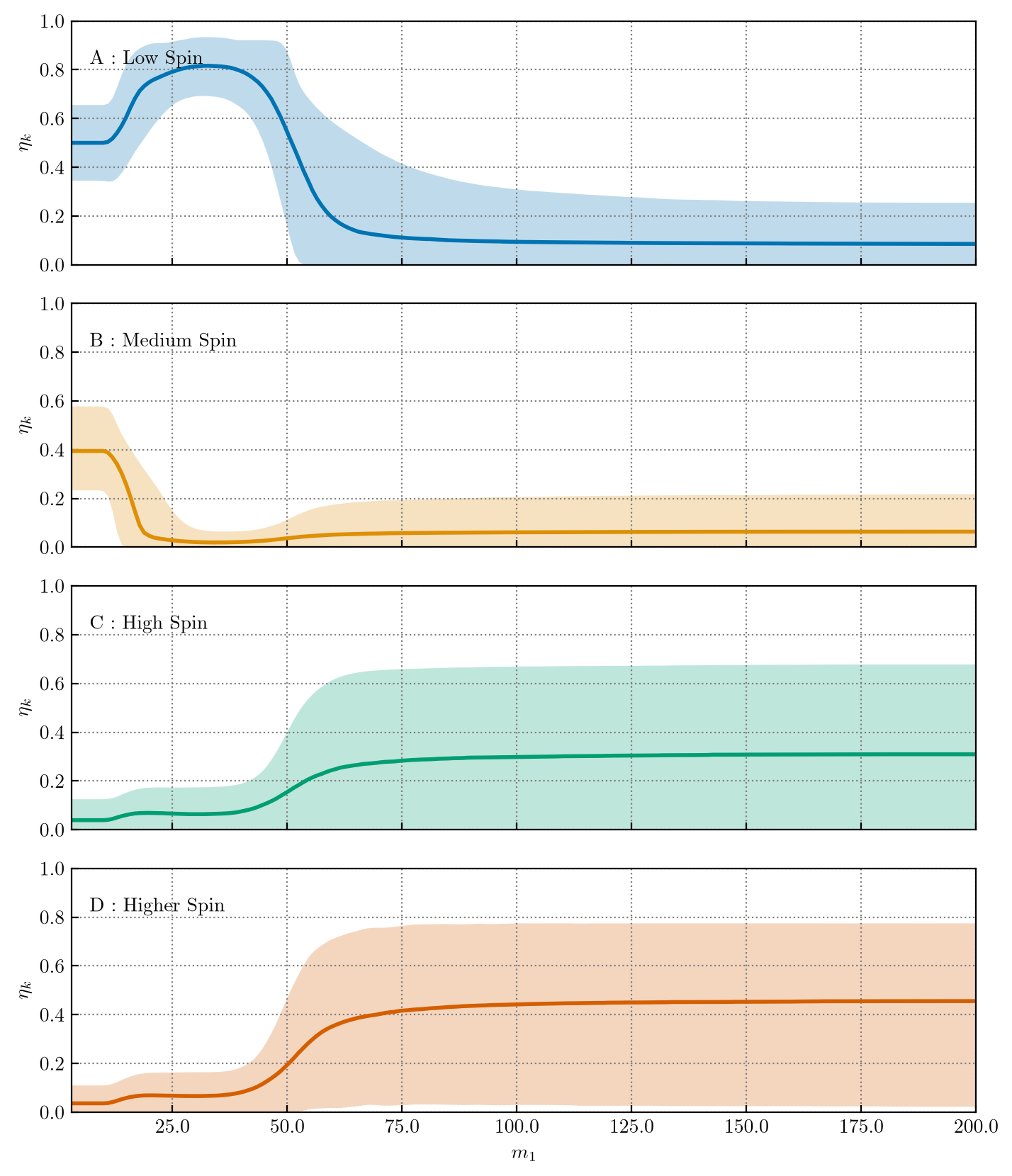}
    \caption{Inferred \acp{PPD} of the mixing fractions when allowing for a fourth subpopulation with mixing fraction $\eta^D(m_1)$. With the additional population, the mixing fractions of the two highest spin populations mirror each other, serving to divide up the probability between the two.}
    \label{fig:four_components_fractions}
\end{figure*}

We do not find evidence for a fourth subpopulation when applying our model to GWTC-5.
We allow for a fourth population with mixing fraction $\eta_D(m_1)$ and with its own spin and mass ratio distribution.
As with our fiducial model, we do not specify any requirements on the labels during sampling, but afterward sort them by the mean of the location parameters of the spin magnitudes.
The results are shown in Figures~\ref{fig:four_components_spins} and~\ref{fig:four_components_fractions}.
The fourth population does not emerge as a distinct population in parameter space. 
In particular, it does not concentrate as a new high-spin population at lower masses $15\,M_\odot \lesssim m_1 \lesssim 25 M_\odot$.
Instead it is confused between the medium- and high-spin populations of the fiducial analysis.

We note that this confusion is not due to our chosen labeling and sorting scheme; we have explored a number of other schemes, such as enforcing labels according to which component dominates at the lowest mass, or by which dominates in regimes where the separation between components is well measured.
In all cases we do not find a distinct fourth population using our model assumptions and dataset.

\subsection{Broader hyperpriors for mass-ratio distribution}

\begin{figure*}
    \centering
    \includegraphics[width=0.48\textwidth]{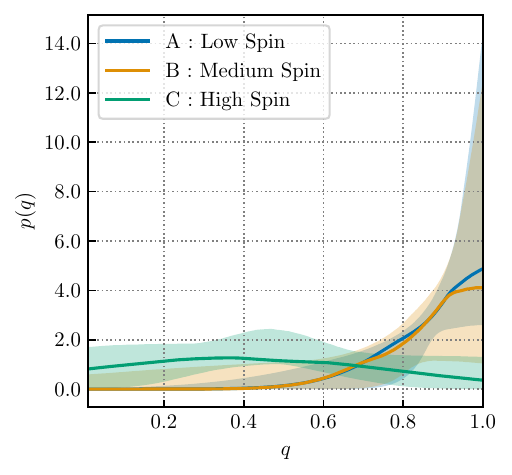}
    \hfill
    \includegraphics[width=0.5\textwidth]{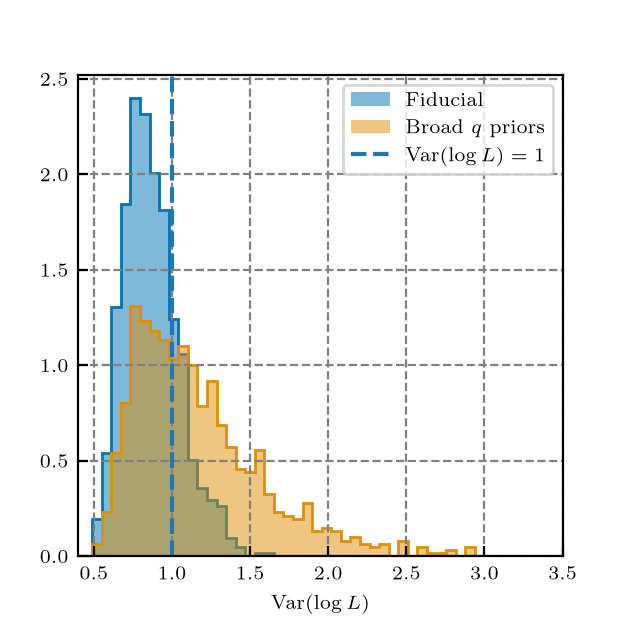}
    \caption{Results from our fiducial model when wider priors are adopted for the hyperparameters $\mu_q^k$ and $\sigma_q^k$ controlling the mass-ratio distribution.
    {\it Left}: The mass-ratio \acp{PPD} for the three populations. As compared to our fiducial priors, the high-spin population shifts slightly towards lower mass ratios (more unequal masses) and the other two populations are allowed to peak more sharply at equal masses.
    {\it Right}: The estimated variance of the log-likelihood evaluated across the hyperposteriors for the two cases. Allowing a greater range of mass ratio distributions increases the variance.}
    \label{fig:Broad_Mass_Ratio_comparison}
\end{figure*}

We relax the bounds on our fiducial model's hyperpriors to allow for narrower mass-ratio distributions that also explore more unequal mass ratios. 
Specifically, we allow $\mu_q \in [0,1.2], \sigma_q \in [0.1,0.8]$ as opposed to the ranges $\mu_q \in [0.4,1.2], \sigma_q \in [0.2,0.8]$ of the fiducial model given in Table~\ref{tab:hyperpriors}.
Our overall results are robust to this change, but when widening the priors we find that the high-spin population shows a greater preference for more unequal masses, as seen in Figure~\ref{fig:Broad_Mass_Ratio_comparison}.
Specifically, we find 
$\mu_q^C \approx \JSONParseValue{\BroadQMacros}{subpop_hyperposteriors.c.mu.mass_ratio}$ 
and  
$\sigma_q^C \approx \JSONParseValue{\BroadQMacros}{subpop_hyperposteriors.c.sigma.mass_ratio}$ 
for the wider priors, as compared to 
$\mu_q^C \approx \JSONParseValue{\FiducialMacros}{subpop_hyperposteriors.c.mu.mass_ratio}$
and $\sigma_q^C \approx \JSONParseValue{\FiducialMacros}{subpop_hyperposteriors.c.sigma.mass_ratio}$ for our fiducial model.
The mass ratio \acp{PPD} for the other two components also display sharper peaks towards equal masses as compared to our fiducial model, with the largest change being in the width of the low-spin population, where $\sigma_q^A \approx \JSONParseValue{\BroadQMacros}{subpop_hyperposteriors.a.sigma.mass_ratio}$ with the wider priors as compared to $\sigma_q^A \approx \JSONParseValue{\FiducialMacros}{subpop_hyperposteriors.a.sigma.mass_ratio}$ for the fiducial model.

The increase in our prior range comes with a cost, namely increased variance in our population likelihood estimates.
Figure~\ref{fig:Broad_Mass_Ratio_comparison} also shows the computed variance of the log-likelihood for both our fiducial model and our model with wider $q$-priors, across our hyperposterior samples.
We see that a fraction of samples from the fiducial run exceed the variance cut of unity applied in~\cite{LIGOScientific:2026ctl} by small amounts.
With the increased prior range, a larger fraction of samples exceeds this variance threshold.

\begin{figure*}
    \centering
    \includegraphics[width=\textwidth]{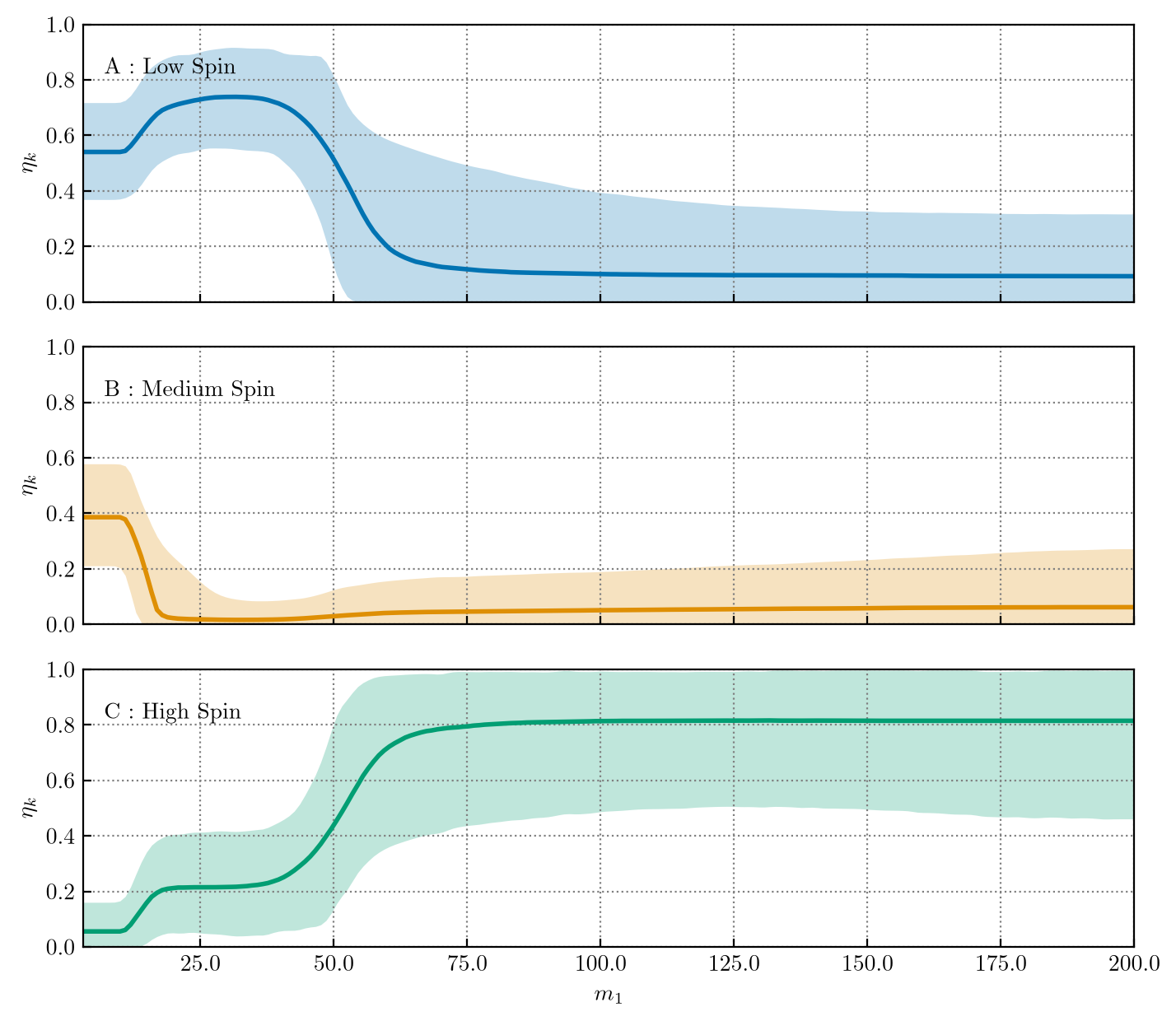}
    \caption{Inferred \acp{PPD} of the mixing fractions when wider priors are adopted for the hyperparameters $\mu_q^k$ and $\sigma_q^k$ controlling the mass-ratio distribution. The high spin population's mixing fraction becomes more uncertain and is elevated at moderate masses ($15\, M_\odot \lesssim m_1 \lesssim 40\, M_\odot$)}
    \label{fig:Broad_Mass_Ratio_fraction}
\end{figure*}

\begin{figure*}
    \centering
    \includegraphics[width=0.48\textwidth]{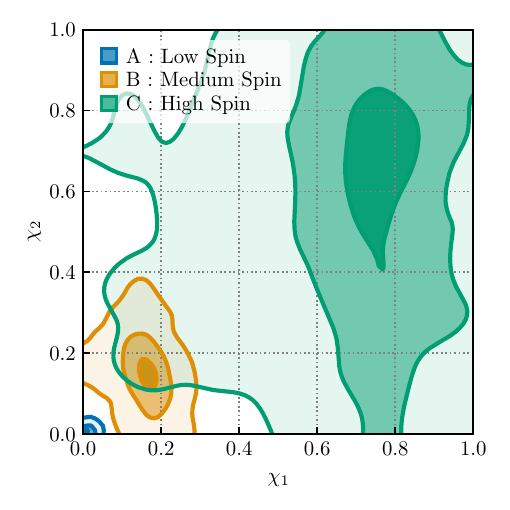}
    \hfill
    \includegraphics[width=0.48\textwidth]{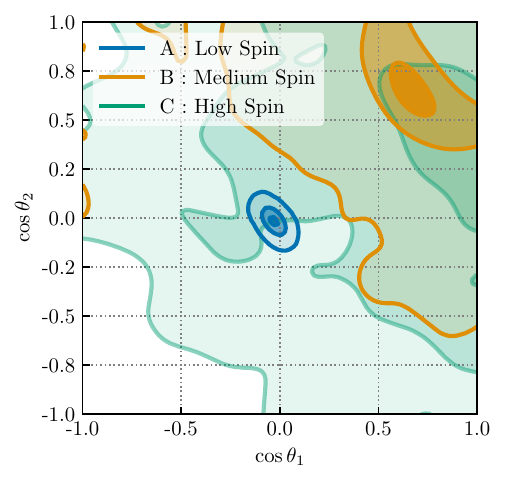}
    \caption{\acp{PPD} over the spin-magnitude (Left) and spin-tilt (Right) when relaxing the fiducial model's bounds on the mass ratio distribution. 
    The high-spin spin-magnitude \ac{PPD} spreads into the region occupied by the medium-spin subpopulation, resulting in mild confusion between the two populations in spin space.}
    \label{fig:Broad_Mass_Ratio_spin_ppds}
\end{figure*}

When we broaden our mass-ratio hyperparameter priors we also introduce greater uncertainty in the high-spin population's mixing fraction $\eta_C(m_1)$, as seen in Figure~\ref{fig:Broad_Mass_Ratio_fraction}.
This fraction is also systematically larger at moderate masses $15\, M_\odot \lesssim m_1 \lesssim 40\, M_\odot$.
The spin-magnitude \ac{PPD} of the high spin population also broadens, leading to greater overlap with the region where the medium-spin population dominates in our fiducial model.
This leads to mild confusion between the high-spin and medium-spin populations in a small region of parameter space at moderate spins, as seen in the spin-magitude and spin-tilt \acp{PPD} in Figure~\ref{fig:Broad_Mass_Ratio_spin_ppds}.

\section{Population assignments of individual binary systems}
\label{sec:assignments}

In this section we provide the assignment probabilities of each source binary for the candidates in GWTC-5 passing our selection threshold and used in our dataset.
As expected, the vast majority lie within the low-spin subpopulation.
Roughly equal numbers of observed binaries lie in each of the medium- and high-spin populations.
As expected, within the high-spin population we find both lower mass but probably rapidly spinning systems like GW241011 and GW241110~\citep{LIGOScientific:2025brd} with high probability, in addition to the highest mass binaries.
Even some of these high mass systems have an appreciable probability of lying within the low-spin subpopulation, which may make up a fraction of the highest mass \acp{BH} within our uncertainties.
For example, GW190521~\citep{LIGOScientific:2020iuh} has greater than 25\% probability of being in the low-spin population but is most probably in the high-spin population.
Meanwhile GW231123~\citep{LIGOScientific:2025rsn} has a posterior assignment probability close to unity for the high-spin population.

\clearpage
\begin{figure}[p]
    \centering
    \includegraphics[width=0.9\textwidth]{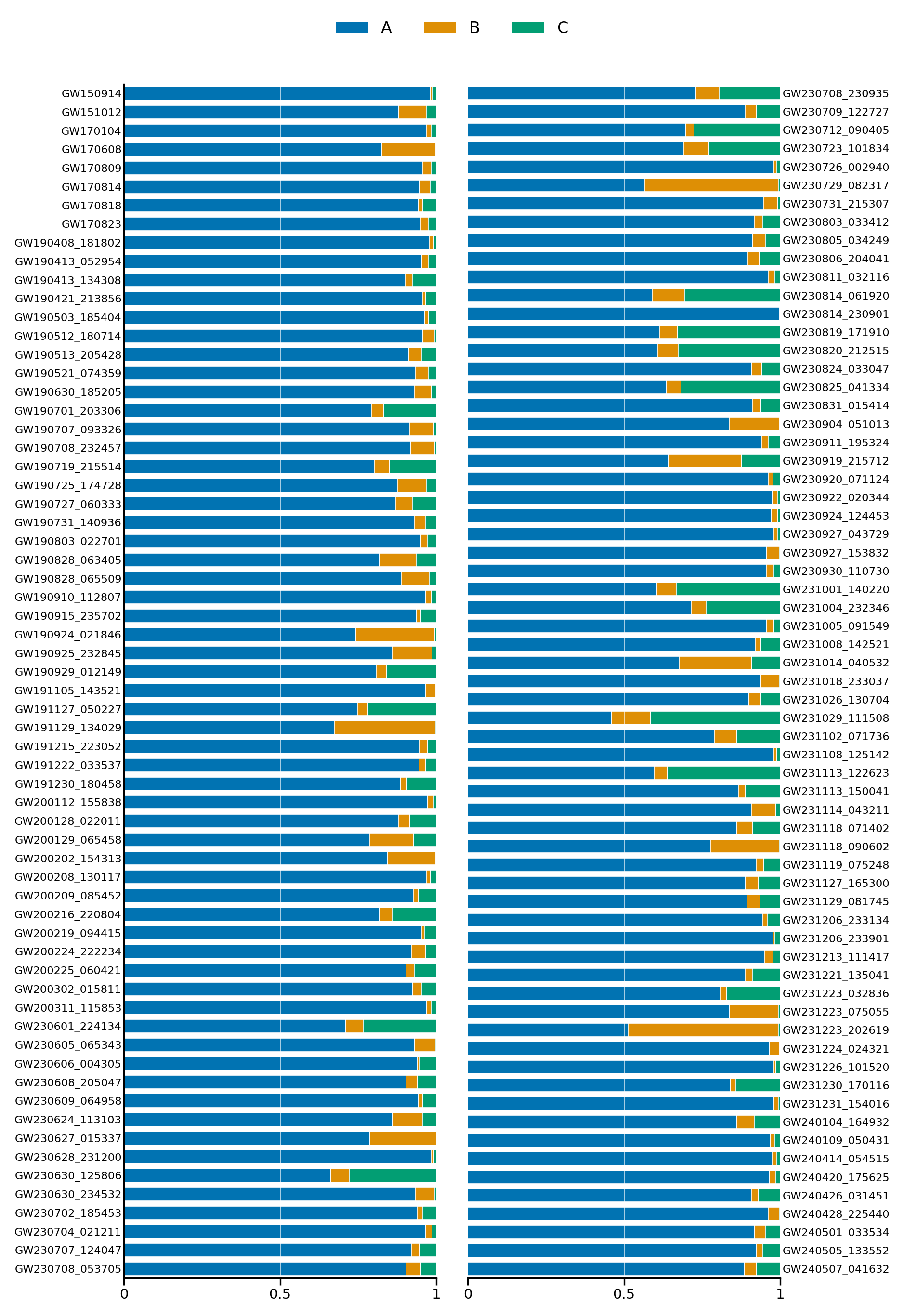}
    \caption{Per-event component-assignment ratios for the first half of the events.}
    \label{fig:event_assignment_ratios_page1}
\end{figure}

\begin{figure}[p]
    \centering
    \includegraphics[width=0.9\textwidth]{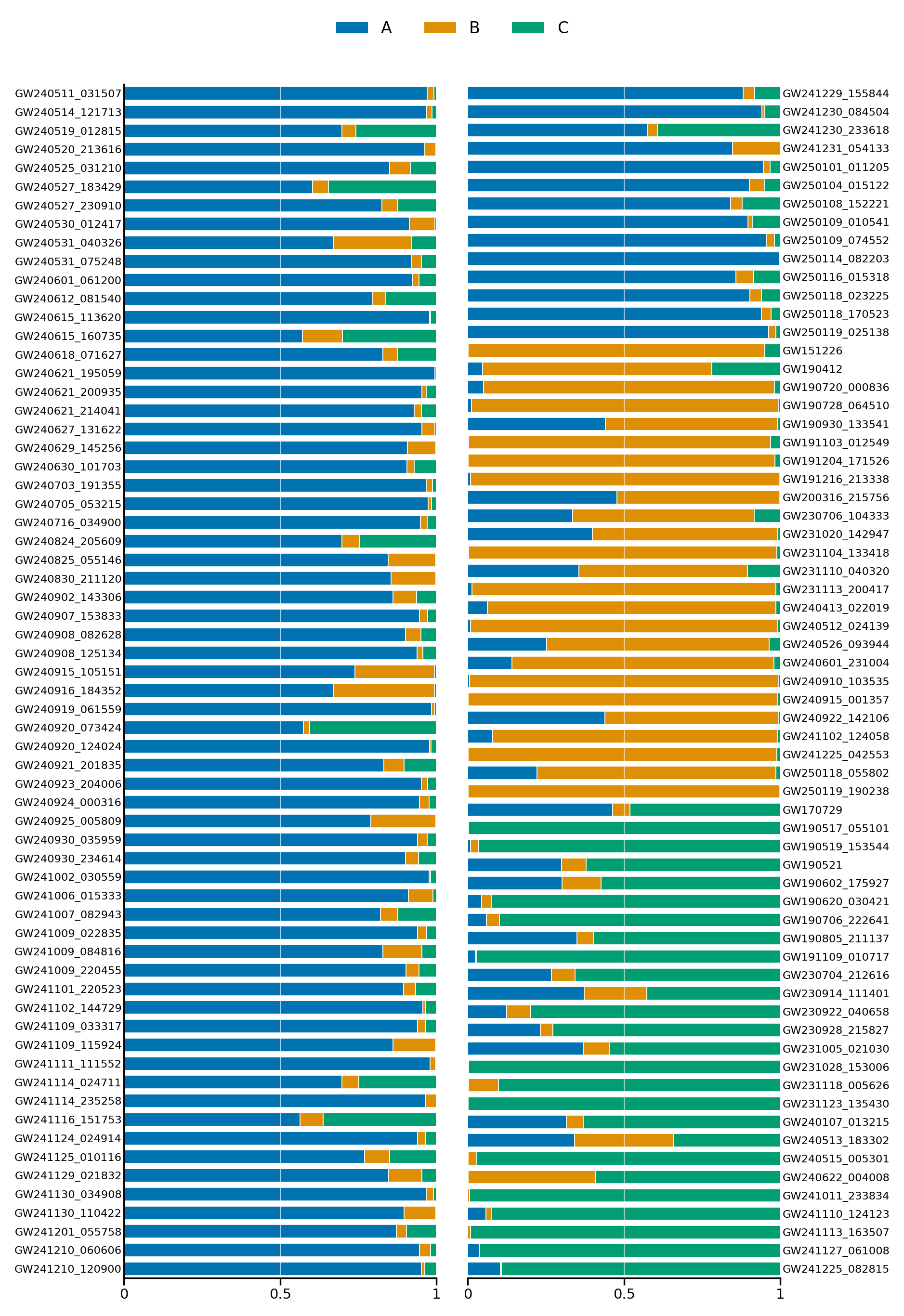}
    \caption{Per-event component-assignment ratios for the second half of the events.}
    \label{fig:event_assignment_ratios_page2}
\end{figure}

\bibliographystyle{aasjournal}
\bibliography{refs.bib}

\begin{thebibliography}{}
\expandafter\ifx\csname natexlab\endcsname\relax\def\natexlab#1{#1}\fi
\providecommand{\url}[1]{\href{#1}{#1}}
\providecommand{\dodoi}[1]{doi:~\href{http://doi.org/#1}{\nolinkurl{#1}}}
\providecommand{\doeprint}[1]{\href{http://ascl.net/#1}{\nolinkurl{http://ascl.net/#1}}}
\providecommand{\doarXiv}[1]{\href{https://arxiv.org/abs/#1}{\nolinkurl{https://arxiv.org/abs/#1}}}

\bibitem[{Aasi {et~al.}(2015)}]{LIGOScientific:2014pky}
Aasi, J., {et~al.} 2015, Class. Quant. Grav., 32, 074001,
  \dodoi{10.1088/0264-9381/32/7/074001}

\bibitem[{Abac {et~al.}(2025{\natexlab{a}})}]{LIGOScientific:2025slb}
Abac, A.~G., {et~al.} 2025{\natexlab{a}}.
\newblock \doarXiv{2508.18082}

\bibitem[{Abac {et~al.}(2025{\natexlab{b}})}]{LIGOScientific:2025pvj}
---. 2025{\natexlab{b}}.
\newblock \doarXiv{2508.18083}

\bibitem[{Abac {et~al.}(2025{\natexlab{c}})}]{LIGOScientific:2025brd}
---. 2025{\natexlab{c}}, Astrophys. J. Lett., 993, L21,
  \dodoi{10.3847/2041-8213/ae0d54}

\bibitem[{Abac {et~al.}(2025{\natexlab{d}})}]{LIGOScientific:2025rsn}
---. 2025{\natexlab{d}}, Astrophys. J. Lett., 993, L25,
  \dodoi{10.3847/2041-8213/ae0c9c}

\bibitem[{Abac {et~al.}(2026{\natexlab{a}})}]{LIGOScientific:2026wfs}
---. 2026{\natexlab{a}}.
\newblock \doarXiv{2605.27225}

\bibitem[{Abac {et~al.}(2026{\natexlab{b}})}]{LIGOScientific:2026ctl}
---. 2026{\natexlab{b}}.
\newblock \doarXiv{2605.27226}

\bibitem[{Abac {et~al.}(2026{\natexlab{c}})}]{LIGOScientific:2026sit}
Abac, N., {et~al.} 2026{\natexlab{c}}.
\newblock \doarXiv{2605.27223}

\bibitem[{Abbott {et~al.}(2020)}]{LIGOScientific:2020iuh}
Abbott, R., {et~al.} 2020, Phys. Rev. Lett., 125, 101102,
  \dodoi{10.1103/PhysRevLett.125.101102}

\bibitem[{Abbott {et~al.}(2023{\natexlab{a}})}]{KAGRA:2021vkt}
---. 2023{\natexlab{a}}, Phys. Rev. X, 13, 041039,
  \dodoi{10.1103/PhysRevX.13.041039}

\bibitem[{Abbott {et~al.}(2023{\natexlab{b}})}]{KAGRA:2021duu}
---. 2023{\natexlab{b}}, Phys. Rev. X, 13, 011048,
  \dodoi{10.1103/PhysRevX.13.011048}

\bibitem[{Acernese {et~al.}(2015)}]{VIRGO:2014yos}
Acernese, F., {et~al.} 2015, Class. Quant. Grav., 32, 024001,
  \dodoi{10.1088/0264-9381/32/2/024001}

\bibitem[{Adamcewicz {et~al.}(2025)Adamcewicz, Guttman, Lasky, \&
  Thrane}]{Adamcewicz:2025phm}
Adamcewicz, C., Guttman, N., Lasky, P.~D., \& Thrane, E. 2025, Astrophys. J.,
  994, 261, \dodoi{10.3847/1538-4357/ae1370}

\bibitem[{Alvarez-Lopez {et~al.}(2026)Alvarez-Lopez, Heinzel, \&
  Vitale}]{Alvarez-Lopez:2026ymo}
Alvarez-Lopez, S., Heinzel, J., \& Vitale, S. 2026.
\newblock \doarXiv{2606.12205}

\bibitem[{Antonini {et~al.}(2025{\natexlab{a}})Antonini, Callister, Dosopoulou,
  Romero-Shaw, \& Chattopadhyay}]{Antonini:2025zzw}
Antonini, F., Callister, T., Dosopoulou, F., Romero-Shaw, I.~M., \&
  Chattopadhyay, D. 2025{\natexlab{a}}, Phys. Rev. D, 112, 063040,
  \dodoi{10.1103/nxnr-pdyx}

\bibitem[{Antonini {et~al.}(2023)Antonini, Gieles, Dosopoulou, \&
  Chattopadhyay}]{Antonini:2022vib}
Antonini, F., Gieles, M., Dosopoulou, F., \& Chattopadhyay, D. 2023, Mon. Not.
  Roy. Astron. Soc., 522, 466, \dodoi{10.1093/mnras/stad972}

\bibitem[{Antonini {et~al.}(2025{\natexlab{b}})Antonini, Romero-Shaw,
  Callister, Dosopoulou, Chattopadhyay, Gieles, \& Mapelli}]{Antonini:2025ilj}
Antonini, F., Romero-Shaw, I., Callister, T., {et~al.} 2025{\natexlab{b}}.
\newblock \doarXiv{2509.04637}

\bibitem[{Antonini {et~al.}(2025{\natexlab{c}})Antonini, Romero-Shaw, \&
  Callister}]{Antonini:2024het}
Antonini, F., Romero-Shaw, I.~M., \& Callister, T. 2025{\natexlab{c}}, Phys.
  Rev. Lett., 134, 011401, \dodoi{10.1103/PhysRevLett.134.011401}

\bibitem[{Antonini {et~al.}(2017)Antonini, Toonen, \&
  Hamers}]{Antonini:2017ash}
Antonini, F., Toonen, S., \& Hamers, A.~S. 2017, Astrophys. J., 841, 77,
  \dodoi{10.3847/1538-4357/aa6f5e}

\bibitem[{Banagiri {et~al.}(2025)Banagiri, Thrane, \& Lasky}]{Banagiri:2025dmy}
Banagiri, S., Thrane, E., \& Lasky, P.~D. 2025.
\newblock \doarXiv{2509.15646}

\bibitem[{Barkat {et~al.}(1967)Barkat, Rakavy, \& Sack}]{Barkat:1967zz}
Barkat, Z., Rakavy, G., \& Sack, N. 1967, Phys. Rev. Lett., 18, 379,
  \dodoi{10.1103/PhysRevLett.18.379}

\bibitem[{Bartos \& Haiman(2026{\natexlab{a}})}]{Bartos:2025pkv}
Bartos, I., \& Haiman, Z. 2026{\natexlab{a}}, Astrophys. J. Lett., 996, L44,
  \dodoi{10.3847/2041-8213/ae2bff}

\bibitem[{Bartos \& Haiman(2026{\natexlab{b}})}]{Bartos:2026xlt}
---. 2026{\natexlab{b}}.
\newblock \doarXiv{2605.09351}

\bibitem[{Bavera {et~al.}(2020)Bavera, Fragos, Qin, Zapartas, Neijssel, Mandel,
  Batta, Gaebel, Kimball, \& Stevenson}]{Bavera:2020inc}
Bavera, S.~S., Fragos, T., Qin, Y., {et~al.} 2020, Astron. Astrophys., 635,
  A97, \dodoi{10.1051/0004-6361/201936204}

\bibitem[{Belczynski {et~al.}(2020)}]{Belczynski:2017gds}
Belczynski, K., {et~al.} 2020, Astron. Astrophys., 636, A104,
  \dodoi{10.1051/0004-6361/201936528}

\bibitem[{Berti {et~al.}(2026)Berti, Crescimbeni, Franciolini, Mastrogiovanni,
  Pani, \& Pierra}]{Berti:2025usa}
Berti, E., Crescimbeni, F., Franciolini, G., {et~al.} 2026, Phys. Rev. D, 113,
  043048, \dodoi{10.1103/3mb7-vnft}

\bibitem[{Besançon {et~al.}(2021)Besançon, Papamarkou, Anthoff, Arslan,
  Byrne, Lin, \& Pearson}]{JSSv098i16}
Besançon, M., Papamarkou, T., Anthoff, D., {et~al.} 2021, Journal of
  Statistical Software, 98, 1, \dodoi{10.18637/jss.v098.i16}

\bibitem[{{Betancourt}(2017)}]{2017arXiv170102434B}
{Betancourt}, M. 2017, arXiv e-prints, arXiv:1701.02434,
  \dodoi{10.48550/arXiv.1701.02434}

\bibitem[{Bezanson {et~al.}(2017)Bezanson, Edelman, Karpinski, \&
  Shah}]{Julia-2017}
Bezanson, J., Edelman, A., Karpinski, S., \& Shah, V.~B. 2017, SIAM {R}eview,
  59, 65, \dodoi{10.1137/141000671}

\bibitem[{Bingham {et~al.}(2019)Bingham, Chen, Jankowiak, Obermeyer, Pradhan,
  Karaletsos, Singh, Szerlip, Horsfall, \& Goodman}]{bingham2019pyro}
Bingham, E., Chen, J.~P., Jankowiak, M., {et~al.} 2019, J. Mach. Learn. Res.,
  20, 28:1.
\newblock \url{http://jmlr.org/papers/v20/18-403.html}

\bibitem[{Bouchet-Valat \& Kamiński(2023)}]{JSSv107i04}
Bouchet-Valat, M., \& Kamiński, B. 2023, Journal of Statistical Software, 107,
  1, \dodoi{10.18637/jss.v107.i04}

\bibitem[{Broekgaarden {et~al.}(2022)Broekgaarden, Stevenson, \&
  Thrane}]{Broekgaarden:2022nst}
Broekgaarden, F.~S., Stevenson, S., \& Thrane, E. 2022, Astrophys. J., 938, 45,
  \dodoi{10.3847/1538-4357/ac8879}

\bibitem[{Callister(2021)}]{Callister:2021gxf}
Callister, T. 2021.
\newblock \doarXiv{2104.09508}

\bibitem[{Callister {et~al.}(2022)Callister, Miller, Chatziioannou, \&
  Farr}]{Callister:2022qwb}
Callister, T.~A., Miller, S.~J., Chatziioannou, K., \& Farr, W.~M. 2022,
  Astrophys. J. Lett., 937, L13, \dodoi{10.3847/2041-8213/ac847e}

\bibitem[{Chattopadhyay {et~al.}(2022)Chattopadhyay, Hurley, Stevenson, \&
  Raidani}]{Chattopadhyay:2022buz}
Chattopadhyay, D., Hurley, J., Stevenson, S., \& Raidani, A. 2022, Mon. Not.
  Roy. Astron. Soc., 513, 4527, \dodoi{10.1093/mnras/stac1163}

\bibitem[{Cheng {et~al.}(2026)Cheng, Toubiana, Biscoveanu, \&
  Gair}]{Cheng:2026bpc}
Cheng, A.~Q., Toubiana, A., Biscoveanu, S., \& Gair, J. 2026.
\newblock \doarXiv{2605.25980}

\bibitem[{Croon {et~al.}(2026)Croon, Gerosa, \& Sakstein}]{Croon:2025gol}
Croon, D., Gerosa, D., \& Sakstein, J. 2026, Mon. Not. Roy. Astron. Soc., 546,
  stag073, \dodoi{10.1093/mnras/stag073}

\bibitem[{De~Luca {et~al.}(2021)De~Luca, Desjacques, Franciolini, Pani, \&
  Riotto}]{DeLuca:2020sae}
De~Luca, V., Desjacques, V., Franciolini, G., Pani, P., \& Riotto, A. 2021,
  Phys. Rev. Lett., 126, 051101, \dodoi{10.1103/PhysRevLett.126.051101}

\bibitem[{De~Luca {et~al.}(2026)De~Luca, Franciolini, \&
  Riotto}]{DeLuca:2025fln}
De~Luca, V., Franciolini, G., \& Riotto, A. 2026, Phys. Rev. Lett., 136,
  201401, \dodoi{10.1103/y4x3-v6j2}

\bibitem[{de~Mink \& Mandel(2016)}]{deMink:2016vkw}
de~Mink, S.~E., \& Mandel, I. 2016, Mon. Not. Roy. Astron. Soc., 460, 3545,
  \dodoi{10.1093/mnras/stw1219}

\bibitem[{Di~Carlo {et~al.}(2019)Di~Carlo, Giacobbo, Mapelli, Pasquato, Spera,
  Wang, \& Haardt}]{DiCarlo:2019pmf}
Di~Carlo, U.~N., Giacobbo, N., Mapelli, M., {et~al.} 2019, Mon. Not. Roy.
  Astron. Soc., 487, 2947, \dodoi{10.1093/mnras/stz1453}

\bibitem[{Dickey(1971)}]{dickey1971weighted}
Dickey, J.~M. 1971, The Annals of Mathematical Statistics, 204

\bibitem[{Dossena {et~al.}(2026)Dossena, Gerosa, \& Bruel}]{Dossena:2026cwf}
Dossena, S., Gerosa, D., \& Bruel, T. 2026.
\newblock \doarXiv{2609.31879}

\bibitem[{Essick(2023)}]{Essick:2023toz}
Essick, R. 2023, Phys. Rev. D, 108, 043011, \dodoi{10.1103/PhysRevD.108.043011}

\bibitem[{Essick \& Farr(2022)}]{Essick:2022ojx}
Essick, R., \& Farr, W. 2022.
\newblock \doarXiv{2204.00461}

\bibitem[{Essick {et~al.}(2025)}]{Essick:2025zed}
Essick, R., {et~al.} 2025, Phys. Rev. D, 112, 102001, \dodoi{10.1103/44x3-hv3y}

\bibitem[{Farmer {et~al.}(2019)Farmer, Renzo, de~Mink, Marchant, \&
  Justham}]{Farmer:2019jed}
Farmer, R., Renzo, M., de~Mink, S.~E., Marchant, P., \& Justham, S. 2019,
  Astrophys. J., 887, 53, \dodoi{10.3847/1538-4357/ab518b}

\bibitem[{Farr(2019)}]{Farr:2019rap}
Farr, W.~M. 2019, Research Notes of the AAS, 3, 66,
  \dodoi{10.3847/2515-5172/ab1d5f}

\bibitem[{Flanagan {et~al.}(2026{\natexlab{a}})Flanagan, Antonini, Callister,
  Chattopadhyay, Dosopoulou, Romero-Shaw, \& Stegmann}]{Flanagan:2026ayy}
Flanagan, E., Antonini, F., Callister, T., {et~al.} 2026{\natexlab{a}}.
\newblock \doarXiv{2606.14472}

\bibitem[{Flanagan {et~al.}(2026{\natexlab{b}})Flanagan, Stegmann, Romero-Shaw,
  Callister, Olejak, \& Antonini}]{Flanagan:2026btu}
Flanagan, E., Stegmann, J., Romero-Shaw, I., {et~al.} 2026{\natexlab{b}}.
\newblock \doarXiv{2607.00565}

\bibitem[{Fowler \& Hoyle(1964)}]{Fowler:1964zz}
Fowler, W.~A., \& Hoyle, F. 1964, Astrophys. J. Suppl., 9, 201,
  \dodoi{10.1086/190103}

\bibitem[{Fuller \& Ma(2019)}]{Fuller:2019sxi}
Fuller, J., \& Ma, L. 2019, Astrophys. J. Lett., 881, L1,
  \dodoi{10.3847/2041-8213/ab339b}

\bibitem[{Fuller {et~al.}(2019)Fuller, Piro, \& Jermyn}]{Fuller:2019ckz}
Fuller, J., Piro, A.~L., \& Jermyn, A.~S. 2019, Mon. Not. Roy. Astron. Soc.,
  485, 3661, \dodoi{10.1093/mnras/stz514}

\bibitem[{Galaudage {et~al.}(2021)Galaudage, Talbot, Nagar, Jain, Thrane, \&
  Mandel}]{Galaudage2021}
Galaudage, S., Talbot, C., Nagar, T., {et~al.} 2021, The Astrophysical Journal
  Letters, 921, L15, \dodoi{10.3847/2041-8213/ac2f3c}

\bibitem[{Gerosa \& Berti(2017)}]{Gerosa:2017kvu}
Gerosa, D., \& Berti, E. 2017, Phys. Rev. D, 95, 124046,
  \dodoi{10.1103/PhysRevD.95.124046}

\bibitem[{Gerosa \& Fishbach(2021)}]{Gerosa:2021mno}
Gerosa, D., \& Fishbach, M. 2021, Nature Astron., 5, 749,
  \dodoi{10.1038/s41550-021-01398-w}

\bibitem[{Gottlieb {et~al.}(2025)Gottlieb, Metzger, Issa, Li, Renzo, \&
  Isi}]{Gottlieb:2025ugy}
Gottlieb, O., Metzger, B.~D., Issa, D., {et~al.} 2025, Astrophys. J. Lett.,
  993, L54, \dodoi{10.3847/2041-8213/ae0d81}

\bibitem[{Guttman {et~al.}(2026{\natexlab{a}})Guttman, Lasky, \&
  Thrane}]{Guttman:2026cnv}
Guttman, N., Lasky, P.~D., \& Thrane, E. 2026{\natexlab{a}}.
\newblock \doarXiv{2607.22011}

\bibitem[{Guttman {et~al.}(2026{\natexlab{b}})Guttman, Payne, Lasky, \&
  Thrane}]{Guttman:2025jkv}
Guttman, N., Payne, E., Lasky, P.~D., \& Thrane, E. 2026{\natexlab{b}},
  Astrophys. J., 996, 144, \dodoi{10.3847/1538-4357/ae17af}

\bibitem[{Harris {et~al.}(2020)}]{numpy}
Harris, C.~R., {et~al.} 2020, Nature, 585, 357,
  \dodoi{10.1038/s41586-020-2649-2}

\bibitem[{Heinzel \& Vitale(2025)}]{Heinzel:2025ogf}
Heinzel, J., \& Vitale, S. 2025.
\newblock \doarXiv{2509.07221}

\bibitem[{{Hoffman} \& {Gelman}(2011)}]{2011arXiv1111.4246H}
{Hoffman}, M.~D., \& {Gelman}, A. 2011, arXiv e-prints, arXiv:1111.4246,
  \dodoi{10.48550/arXiv.1111.4246}

\bibitem[{{Hoy} {et~al.}(2022){Hoy}, {Fairhurst}, {Hannam}, \&
  {Tiwari}}]{2022ApJ...928...75H}
{Hoy}, C., {Fairhurst}, S., {Hannam}, M., \& {Tiwari}, V. 2022, \apj, 928, 75,
  \dodoi{10.3847/1538-4357/ac54a3}

\bibitem[{Hunter(2007)}]{matplotlib}
Hunter, J.~D. 2007, Computing in Science \& Engineering, 9, 90,
  \dodoi{10.1109/MCSE.2007.55}

\bibitem[{Hussain(2024{\natexlab{a}})}]{truncatedgaussianmixtures}
Hussain, A. 2024{\natexlab{a}},
  \href{https://github.com/potatoasad/truncatedgaussianmixtures}{\texttt{truncatedgaussianmixtures}}:
  fitting truncated gaussian mixture models,  Zenodo,
  \dodoi{10.5281/ZENODO.13999561}

\bibitem[{Hussain(2024{\natexlab{b}})}]{gravpop}
---. 2024{\natexlab{b}},
  \href{https://git.ligo.org/asad.hussain/gravpop}{\texttt{gravpop}}:
  Astrophysical population modelling for gravitational waves with the ability
  to probe narrow population features over bounded domains.,  Zenodo,
  \dodoi{10.5281/ZENODO.14003052}

\bibitem[{Hussain {et~al.}(2025)Hussain, Isi, \& Zimmerman}]{Hussain:2025llf}
Hussain, A., Isi, M., \& Zimmerman, A. 2025.
\newblock \doarXiv{2510.20010}

\bibitem[{Hussain {et~al.}(2026{\natexlab{a}})Hussain, Isi, \&
  Zimmerman}]{Hussain2025}
---. 2026{\natexlab{a}}, Astrophys. J., 996, 71

\bibitem[{Hussain {et~al.}(2026{\natexlab{b}})Hussain, Isi, \&
  Zimmerman}]{Hussain:2026pfm}
---. 2026{\natexlab{b}}.
\newblock \doarXiv{2605.24281}

\bibitem[{Hussain {et~al.}(2026{\natexlab{c}})Hussain, Isi, \&
  Zimmerman}]{Hussain:2026data}
---. 2026{\natexlab{c}}, Data release for ``Three subpopulations of binary
  black holes in mass, spin magnitudes and tilts following GWTC-5'', GitHub
  repository.
\newblock \url{https://github.com/Potatoasad/UnravelingFormationPathwaysGWTC5}

\bibitem[{Jeffreys(1939)}]{Jeffreys1939-JEFTOP-5}
Jeffreys, H. 1939, Theory of Probability (Oxford, England: Clarendon Press)

\bibitem[{Johnson(2013)}]{quadgk}
Johnson, S.~G. 2013, {QuadGK.jl}: {G}auss--{K}ronrod integration in {J}ulia,
  \url{https://github.com/JuliaMath/QuadGK.jl}

\bibitem[{Kass \& Raftery(1995)}]{Kass01061995}
Kass, R.~E., \& Raftery, A.~E. 1995, Journal of the American Statistical
  Association, 90, 773, \dodoi{10.1080/01621459.1995.10476572}

\bibitem[{Kimball {et~al.}(2020)Kimball, Talbot, L.~Berry, Carney, Zevin,
  Thrane, \& Kalogera}]{Kimball:2020opk}
Kimball, C., Talbot, C., L.~Berry, C.~P., {et~al.} 2020, Astrophys. J., 900,
  177, \dodoi{10.3847/1538-4357/aba518}

\bibitem[{K{\i}ro{\u{g}}lu {et~al.}(2025)K{\i}ro{\u{g}}lu, Kremer, \&
  Rasio}]{Kiroglu:2025vqy}
K{\i}ro{\u{g}}lu, F., Kremer, K., \& Rasio, F.~A. 2025, Astrophys. J. Lett.,
  994, L37, \dodoi{10.3847/2041-8213/ae1eeb}

\bibitem[{Kremer {et~al.}(2020)Kremer, Spera, Becker, Chatterjee, Di~Carlo,
  Fragione, Rodriguez, Ye, \& Rasio}]{Kremer:2020wtp}
Kremer, K., Spera, M., Becker, D., {et~al.} 2020, Astrophys. J., 903, 45,
  \dodoi{10.3847/1538-4357/abb945}

\bibitem[{Li {et~al.}(2025)Li, Wang, Tang, \& Fan}]{Li:2025iux}
Li, Y.-J., Wang, Y.-Z., Tang, S.-P., \& Fan, Y.-Z. 2025.
\newblock \doarXiv{2509.23897}

\bibitem[{Li {et~al.}(2026)Li, Wang, Tang, \& Fan}]{Li:2026iae}
---. 2026.
\newblock \doarXiv{2607.01121}

\bibitem[{Lin {et~al.}(2019)Lin, White, Byrne, Bates, Noack, Pearson, Arslan,
  Squire, Anthoff, Papamarkou, Besançon, Drugowitsch, Schauer, \&
  contributors}]{Distributions.jl-2019}
Lin, D., White, J.~M., Byrne, S., {et~al.} 2019, {JuliaStats/Distributions.jl:
  a Julia package for probability distributions and associated functions},
  \dodoi{10.5281/zenodo.2647458}

\bibitem[{Liu \& Lai(2018)}]{Liu:2018nrf}
Liu, B., \& Lai, D. 2018, Astrophys. J., 863, 68,
  \dodoi{10.3847/1538-4357/aad09f}

\bibitem[{{LVK
  Collaboration}(2026)}]{ligo_scientific_collaboration_2026_19500052}
{LVK Collaboration}. 2026, GWTC-5.0: Cumulative Search Sensitivity Estimates,
  Zenodo, \dodoi{10.5281/zenodo.19500052}

\bibitem[{Ma \& Fuller(2019)}]{Ma:2019cpr}
Ma, L., \& Fuller, J. 2019, Mon. Not. Roy. Astron. Soc., 488, 4338,
  \dodoi{10.1093/mnras/stz2009}

\bibitem[{Mandel \& de~Mink(2016)}]{Mandel:2015qlu}
Mandel, I., \& de~Mink, S.~E. 2016, Mon. Not. Roy. Astron. Soc., 458, 2634,
  \dodoi{10.1093/mnras/stw379}

\bibitem[{Mandel \& Farmer(2022)}]{Mandel:2018hfr}
Mandel, I., \& Farmer, A. 2022, Phys. Rept., 955, 1,
  \dodoi{10.1016/j.physrep.2022.01.003}

\bibitem[{Mandel {et~al.}(2019)Mandel, Farr, \& Gair}]{Mandel:2018mve}
Mandel, I., Farr, W.~M., \& Gair, J.~R. 2019, Mon. Not. Roy. Astron. Soc., 486,
  1086, \dodoi{10.1093/mnras/stz896}

\bibitem[{Mapelli(2021)}]{Mapelli:2021taw}
Mapelli, M. 2021, in Handbook of Gravitational Wave Astronomy, ed. C.~{Bambi},
  S.~{Katsanevas}, \& K.~D. {Kokkotas} (Springer, Singapore), 16,
  \dodoi{10.1007/978-981-15-4702-7_16-1}

\bibitem[{Marchant {et~al.}(2016)Marchant, Langer, Podsiadlowski, Tauris, \&
  Moriya}]{Marchant:2016wow}
Marchant, P., Langer, N., Podsiadlowski, P., Tauris, T.~M., \& Moriya, T.~J.
  2016, Astron. Astrophys., 588, A50, \dodoi{10.1051/0004-6361/201628133}

\bibitem[{{McKernan} \& {Ford}(2024)}]{2024MNRAS.531.3479M}
{McKernan}, B., \& {Ford}, K.~E.~S. 2024, \mnras, 531, 3479,
  \dodoi{10.1093/mnras/stae1351}

\bibitem[{McKernan {et~al.}(2020)McKernan, Ford, O'Shaughnessy, \&
  Wysocki}]{McKernan:2019beu}
McKernan, B., Ford, K. E.~S., O'Shaughnessy, R., \& Wysocki, D. 2020, Mon. Not.
  Roy. Astron. Soc., 494, 1203, \dodoi{10.1093/mnras/staa740}

\bibitem[{Mehta {et~al.}(2023)Mehta, Olsen, Wadekar, Roulet, Venumadhav,
  Mushkin, Zackay, \& Zaldarriaga}]{Mehta:2023zlk}
Mehta, A.~K., Olsen, S., Wadekar, D., {et~al.} 2023.
\newblock \doarXiv{2311.06061}

\bibitem[{Mould {et~al.}(2022)Mould, Gerosa, Broekgaarden, \&
  Steinle}]{Mould:2022xeu}
Mould, M., Gerosa, D., Broekgaarden, F.~S., \& Steinle, N. 2022, Mon. Not. Roy.
  Astron. Soc., 517, 2738, \dodoi{10.1093/mnras/stac2859}

\bibitem[{{Neal}(2011)}]{2011hmcm.book..113N}
{Neal}, R. 2011, in Handbook of Markov Chain Monte Carlo (Chapman \& Hall/CRC),
  113--162, \dodoi{10.1201/b10905}

\bibitem[{Nitz {et~al.}(2023)Nitz, Kumar, Wang, Kastha, Wu, Sch\"afer,
  Dhurkunde, \& Capano}]{Nitz:2021zwj}
Nitz, A.~H., Kumar, S., Wang, Y.-F., {et~al.} 2023, Astrophys. J., 946, 59,
  \dodoi{10.3847/1538-4357/aca591}

\bibitem[{Olejak \& Belczynski(2021)}]{Olejak:2021iux}
Olejak, A., \& Belczynski, K. 2021, Astrophys. J. Lett., 921, L2,
  \dodoi{10.3847/2041-8213/ac2f48}

\bibitem[{Olsen {et~al.}(2022)Olsen, Venumadhav, Mushkin, Roulet, Zackay, \&
  Zaldarriaga}]{Olsen:2022pin}
Olsen, S., Venumadhav, T., Mushkin, J., {et~al.} 2022, Phys. Rev. D, 106,
  043009, \dodoi{10.1103/PhysRevD.106.043009}

\bibitem[{Padhyegurjar \& Mukherjee(2026)}]{Padhyegurjar:2026scg}
Padhyegurjar, S., \& Mukherjee, S. 2026.
\newblock \doarXiv{2606.00234}

\bibitem[{pandas~development team(2020)}]{pandas}
pandas~development team, T. 2020, pandas-dev/pandas: Pandas, latest,  Zenodo,
  \dodoi{10.5281/zenodo.3509134}

\bibitem[{Park {et~al.}(2017)Park, Kim, Lee, Bae, \& Belczynski}]{Park:2017zgj}
Park, D., Kim, C., Lee, H.~M., Bae, Y.-B., \& Belczynski, K. 2017, Mon. Not.
  Roy. Astron. Soc., 469, 4665, \dodoi{10.1093/mnras/stx1015}

\bibitem[{Phan {et~al.}(2019)Phan, Pradhan, \& Jankowiak}]{phan2019composable}
Phan, D., Pradhan, N., \& Jankowiak, M. 2019, arXiv preprint arXiv:1912.11554

\bibitem[{Pierra {et~al.}(2024)Pierra, Mastrogiovanni, \&
  Perri{\`e}s}]{Pierra:2024fbl}
Pierra, G., Mastrogiovanni, S., \& Perri{\`e}s, S. 2024, Astron. Astrophys.,
  692, A80, \dodoi{10.1051/0004-6361/202452545}

\bibitem[{Plunkett {et~al.}(2026)Plunkett, Vitale, Callister, \&
  Zevin}]{Plunkett:2026pxt}
Plunkett, C., Vitale, S., Callister, T., \& Zevin, M. 2026, Phys. Rev. Lett.,
  137, 021404, \dodoi{10.1103/n6p4-ftgq}

\bibitem[{Qin {et~al.}(2018)Qin, Fragos, Meynet, Andrews, S\o{}rensen, \&
  Song}]{Qin:2018vaa}
Qin, Y., Fragos, T., Meynet, G., {et~al.} 2018, Astron. Astrophys., 616, A28,
  \dodoi{10.1051/0004-6361/201832839}

\bibitem[{Ray \& Kalogera(2026{\natexlab{a}})}]{Ray:2026fdm}
Ray, A., \& Kalogera, V. 2026{\natexlab{a}}.
\newblock \doarXiv{2607.28622}

\bibitem[{Ray \& Kalogera(2026{\natexlab{b}})}]{Ray:2025xti}
---. 2026{\natexlab{b}}, Astrophys. J. Lett., 998, L20,
  \dodoi{10.3847/2041-8213/ae374d}

\bibitem[{Ray {et~al.}(2025)Ray, Maga{\~n}a~Hernandez, Breivik, \&
  Creighton}]{Ray:2024hos}
Ray, A., Maga{\~n}a~Hernandez, I., Breivik, K., \& Creighton, J. 2025,
  Astrophys. J., 991, 17, \dodoi{10.3847/1538-4357/adf22a}

\bibitem[{Ray {et~al.}(2026)Ray, Mukherjee, Zevin, \& Kalogera}]{Ray:2026uur}
Ray, A., Mukherjee, S., Zevin, M., \& Kalogera, V. 2026.
\newblock \doarXiv{2603.17987}

\bibitem[{Riley {et~al.}(2021)Riley, Mandel, Marchant, Butler, Nathaniel,
  Neijssel, Shortt, \& Vigna-Gomez}]{Riley:2020btf}
Riley, J., Mandel, I., Marchant, P., {et~al.} 2021, Mon. Not. Roy. Astron.
  Soc., 505, 663, \dodoi{10.1093/mnras/stab1291}

\bibitem[{Rodriguez \& Antonini(2018)}]{Rodriguez:2018jqu}
Rodriguez, C.~L., \& Antonini, F. 2018, Astrophys. J., 863, 7,
  \dodoi{10.3847/1538-4357/aacea4}

\bibitem[{Rodriguez {et~al.}(2016{\natexlab{a}})Rodriguez, Haster, Chatterjee,
  Kalogera, \& Rasio}]{Rodriguez:2016avt}
Rodriguez, C.~L., Haster, C.-J., Chatterjee, S., Kalogera, V., \& Rasio, F.~A.
  2016{\natexlab{a}}, Astrophys. J. Lett., 824, L8,
  \dodoi{10.3847/2041-8205/824/1/L8}

\bibitem[{Rodriguez {et~al.}(2019)Rodriguez, Zevin, Amaro-Seoane, Chatterjee,
  Kremer, Rasio, \& Ye}]{Rodriguez:2019huv}
Rodriguez, C.~L., Zevin, M., Amaro-Seoane, P., {et~al.} 2019, Phys. Rev. D,
  100, 043027, \dodoi{10.1103/PhysRevD.100.043027}

\bibitem[{Rodriguez {et~al.}(2016{\natexlab{b}})Rodriguez, Zevin, Pankow,
  Kalogera, \& Rasio}]{Rodriguez:2016vmx}
Rodriguez, C.~L., Zevin, M., Pankow, C., Kalogera, V., \& Rasio, F.~A.
  2016{\natexlab{b}}, Astrophys. J. Lett., 832, L2,
  \dodoi{10.3847/2041-8205/832/1/L2}

\bibitem[{Shibata \& Fujibayashi(2026)}]{Shibata:2025lde}
Shibata, M., \& Fujibayashi, S. 2026, Astrophys. J., 996, 57,
  \dodoi{10.3847/1538-4357/ae22d4}

\bibitem[{Spera {et~al.}(2022)Spera, Trani, \& Mencagli}]{Spera:2022byb}
Spera, M., Trani, A.~A., \& Mencagli, M. 2022, Galaxies, 10, 76,
  \dodoi{10.3390/galaxies10040076}

\bibitem[{Sridhar {et~al.}(2025)Sridhar, Ray, \& Kalogera}]{Sridhar:2025kvi}
Sridhar, O., Ray, A., \& Kalogera, V. 2025.
\newblock \doarXiv{2511.22093}

\bibitem[{Stegmann {et~al.}(2026)Stegmann, Antonini, Olejak, Biscoveanu,
  Raymond, Rinaldi, \& Flanagan}]{Stegmann:2025zkb}
Stegmann, J., Antonini, F., Olejak, A., {et~al.} 2026, Astrophys. J. Lett.,
  1000, L59, \dodoi{10.3847/2041-8213/ae52ec}

\bibitem[{Tagawa {et~al.}(2021)Tagawa, Haiman, Bartos, Kocsis, \&
  Omukai}]{Tagawa2021}
Tagawa, H., Haiman, Z., Bartos, I., Kocsis, B., \& Omukai, K. 2021, Monthly
  Notices of the Royal Astronomical Society, 507, 3362,
  \dodoi{10.1093/mnras/stab2315}

\bibitem[{Talbot \& Golomb(2023)}]{Talbot:2023pex}
Talbot, C., \& Golomb, J. 2023, Mon. Not. Roy. Astron. Soc., 526, 3495,
  \dodoi{10.1093/mnras/stad2968}

\bibitem[{Thrane \& Talbot(2019)}]{Thrane:2018qnx}
Thrane, E., \& Talbot, C. 2019, Publ. Astron. Soc. Austral., 36, e010,
  \dodoi{10.1017/pasa.2019.2}

\bibitem[{Tong {et~al.}(2025)Tong, Callister, Fishbach, Thrane, Antonini,
  Stevenson, Romero-Shaw, \& Dosopoulou}]{Tong:2025xir}
Tong, H., Callister, T.~A., Fishbach, M., {et~al.} 2025.
\newblock \doarXiv{2511.05316}

\bibitem[{Tong {et~al.}(2026)}]{Tong:2025wpz}
Tong, H., {et~al.} 2026, Nature, 652, 874, \dodoi{10.1038/s41586-026-10359-0}

\bibitem[{Uchida {et~al.}(2019)Uchida, Shibata, Takahashi, \&
  Yoshida}]{Uchida:2019gjx}
Uchida, H., Shibata, M., Takahashi, K., \& Yoshida, T. 2019, Phys. Rev. D, 99,
  041302, \dodoi{10.1103/PhysRevD.99.041302}

\bibitem[{Vaccaro {et~al.}(2024)Vaccaro, Mapelli, P{\'e}rigois, Barone, Artale,
  Dall'Amico, Iorio, \& Torniamenti}]{Vaccaro:2023cwr}
Vaccaro, M.~P., Mapelli, M., P{\'e}rigois, C., {et~al.} 2024, Astron.
  Astrophys., 685, A51, \dodoi{10.1051/0004-6361/202348509}

\bibitem[{Van~Rossum \& Drake(2009)}]{Python3}
Van~Rossum, G., \& Drake, F.~L. 2009, Python 3 Reference Manual (Scotts Valley,
  CA: CreateSpace)

\bibitem[{Venumadhav {et~al.}(2019)Venumadhav, Zackay, Roulet, Dai, \&
  Zaldarriaga}]{Venumadhav:2019tad}
Venumadhav, T., Zackay, B., Roulet, J., Dai, L., \& Zaldarriaga, M. 2019, Phys.
  Rev. D, 100, 023011, \dodoi{10.1103/PhysRevD.100.023011}

\bibitem[{Venumadhav {et~al.}(2020)Venumadhav, Zackay, Roulet, Dai, \&
  Zaldarriaga}]{Venumadhav:2019lyq}
---. 2020, Phys. Rev. D, 101, 083030, \dodoi{10.1103/PhysRevD.101.083030}

\bibitem[{Verdinelli \& Wasserman(1995)}]{Verdinelli01061995}
Verdinelli, I., \& Wasserman, L. 1995, Journal of the American Statistical
  Association, 90, 614, \dodoi{10.1080/01621459.1995.10476554}

\bibitem[{Verzani(2020)}]{Roots.jl}
Verzani, J. 2020, {Roots.jl}: Root finding functions for Julia,
  \url{https://github.com/JuliaMath/Roots.jl}

\bibitem[{Vijaykumar {et~al.}(2026)Vijaykumar, Farah, \&
  Fishbach}]{Vijaykumar:2026zjy}
Vijaykumar, A., Farah, A.~M., \& Fishbach, M. 2026, Astrophys. J. Lett., 999,
  L30, \dodoi{10.3847/2041-8213/ae4878}

\bibitem[{Virtanen {et~al.}(2020)}]{scipy}
Virtanen, P., {et~al.} 2020, Nature Methods, 17, 261,
  \dodoi{10.1038/s41592-019-0686-2}

\bibitem[{Vitale {et~al.}(2022{\natexlab{a}})Vitale, Biscoveanu, \&
  Talbot}]{Vitale:2022dpa}
Vitale, S., Biscoveanu, S., \& Talbot, C. 2022{\natexlab{a}}, Astron.
  Astrophys., 668, L2, \dodoi{10.1051/0004-6361/202245084}

\bibitem[{Vitale {et~al.}(2022{\natexlab{b}})Vitale, Gerosa, Farr, \&
  Taylor}]{Vitale:2020aaz}
Vitale, S., Gerosa, D., Farr, W.~M., \& Taylor, S.~R. 2022{\natexlab{b}}, in
  Handbook of Gravitational Wave Astronomy, ed. C.~{Bambi}, S.~{Katsanevas}, \&
  K.~D. {Kokkotas} (Springer, Singapore), 45,
  \dodoi{10.1007/978-981-15-4702-7_45-1}

\bibitem[{Vitale \& Mould(2025)}]{Vitale:2025lms}
Vitale, S., \& Mould, M. 2025, Phys. Rev. D, 112, 083015,
  \dodoi{10.1103/drsl-n3wz}

\bibitem[{Wadekar {et~al.}(2023)Wadekar, Roulet, Venumadhav, Mehta, Zackay,
  Mushkin, Olsen, \& Zaldarriaga}]{Wadekar:2023gea}
Wadekar, D., Roulet, J., Venumadhav, T., {et~al.} 2023.
\newblock \doarXiv{2312.06631}

\bibitem[{Wolfe {et~al.}(2026{\natexlab{a}})Wolfe, Hussain, Heinzel, \&
  Vitale}]{Wolfe:2026pjn}
Wolfe, N.~E., Hussain, A., Heinzel, J., \& Vitale, S. 2026{\natexlab{a}}.
\newblock \doarXiv{2609.10753}

\bibitem[{Wolfe {et~al.}(2026{\natexlab{b}})Wolfe, Vitale, \&
  Zevin}]{Wolfe:2026meb}
Wolfe, N.~E., Vitale, S., \& Zevin, M. 2026{\natexlab{b}}, Astrophys. J. Lett.,
  1007, L45, \dodoi{10.3847/2041-8213/ae90ae}

\bibitem[{Woosley(2017)}]{Woosley:2016hmi}
Woosley, S.~E. 2017, Astrophys. J., 836, 244,
  \dodoi{10.3847/1538-4357/836/2/244}

\bibitem[{Woosley \& Heger(2021)}]{Woosley:2021xba}
Woosley, S.~E., \& Heger, A. 2021, Astrophys. J. Lett., 912, L31,
  \dodoi{10.3847/2041-8213/abf2c4}

\bibitem[{Yang {et~al.}(2019)Yang, Bartos, Gayathri, Ford, Haiman, Klimenko,
  Kocsis, Marka, Marka, McKernan, \& O'Shaughnessy}]{Yang2019}
Yang, Y., Bartos, I., Gayathri, V., {et~al.} 2019, Physical Review Letters,
  123, 181101, \dodoi{10.1103/PhysRevLett.123.181101}

\bibitem[{Zackay {et~al.}(2021)Zackay, Dai, Venumadhav, Roulet, \&
  Zaldarriaga}]{Zackay:2019btq}
Zackay, B., Dai, L., Venumadhav, T., Roulet, J., \& Zaldarriaga, M. 2021, Phys.
  Rev. D, 104, 063030, \dodoi{10.1103/PhysRevD.104.063030}

\bibitem[{Zeeshan \& O'Shaughnessy(2026)}]{Zeeshan:2026mpr}
Zeeshan, M., \& O'Shaughnessy, R. 2026.
\newblock \doarXiv{2607.15431}

\bibitem[{Zevin \& Holz(2022)}]{Zevin:2022bfa}
Zevin, M., \& Holz, D.~E. 2022, Astrophys. J. Lett., 935, L20,
  \dodoi{10.3847/2041-8213/ac853d}

\end{thebibliography}

\end{document}